\documentclass[conference,letterpaper]{IEEEtran}
\IEEEoverridecommandlockouts
\usepackage[utf8]{inputenc} 
\usepackage[T1]{fontenc}
\usepackage{url}
\usepackage{ifthen}
\usepackage{amsthm,amssymb,amsfonts}
\usepackage{algorithmicx}
\usepackage{algorithm}
\usepackage{algpseudocode}
\usepackage{graphicx}
\usepackage{textcomp}
\usepackage{tikz}
\usepackage{comment}
\usepackage{tikz-3dplot}
\usepackage[outline]{contour} 
\usepackage{xcolor}
\usepackage{subfiles}
\usepackage{siunitx}
\usepackage{epsfig}

\newcommand{\tauvec}{\boldsymbol{\tau}} 
\newcommand{\nuvec}{\boldsymbol{\nu}} 
\newcommand{\omegavec}{\boldsymbol{\omega}} 
\newcommand{\omegatilde}{\tilde{\omega}} 
\newcommand{\Exp}{\operatorname{Exp}}
\newcommand{\Log}{\operatorname{Log}}
\newcommand{\diag}{\operatorname{diag}}
\newcommand{\Tws}{\mathbf{T}_{w,s}}
\newcommand{\Tsw}{\mathbf{T}_{s,w}}
\newcommand{\Twp}{\mathbf{T}_{w,p}}
\newcommand{\ysp}{\mathbf{y}_{s,p}}
\newcommand{\hsp}{\mathbf{h}_{s,p}}
\newcommand{\zetasp}{\boldsymbol{\zeta}_{s,p}}
\newcommand{\zetasphat}{\hat{\boldsymbol{\zeta}}_{s,p}}
\newcommand{\Tsp}{\mathbf{T}_{s,p}}
\newcommand{\Tsphat}{\hat{\mathbf{T}}_{s,p}}
\newcommand{\Herm}{\operatorname{H}}
\newcommand{\T}{\mathbf{T}}
\newcommand{\rvec}{\mathbf{r}}

\newcommand{\thetavec}{\boldsymbol{\theta}}
\newcommand{\rhovec}{\boldsymbol{\rho}}
\newcommand{\zero}{\mathbf{0}}
\newcommand{\rnorm}{\|\mathbf{r}\|}
\newcommand{\J}{\mathbf{J}}

\newcommand{\Kmat}{\mathbf{K}}
\newcommand{\Mmat}{\mathbf{M}}
\newcommand{\Rmat}{\mathbf{R}}
\newcommand{\Bmat}{\mathbf{B}}
\newcommand{\Abar}{\bar{\mathbf{A}}}

\newcommand{\Smat}{\mathbf{S}}
\newcommand{\Etilde}{\tilde{\mathbf{E}}}
\newcommand{\Ptilde}{\tilde{\mathbf{P}}}
\newcommand{\efour}{\mathbf{e}_4}
\newcommand{\ethree}{\mathbf{e}_3}
\newcommand{\etwo}{\mathbf{e}_2}
\newcommand{\eone}{\mathbf{e}_1}
\newcommand{\F}{\mathbf{F}}
\newcommand{\vs}{\boldsymbol{\xi}_s}
\newcommand{\bvs}{\bar{\boldsymbol{\xi}}_s}
\newcommand{\I}{\mathbf{I}}

\newcommand{\D}{\operatorname{D}}
\newcommand{\R}{\mathbb{R}}

\newcommand{\nuwB}{^{\mathcal{B}}\boldsymbol{\nu}_\mathbf{w}}

\newcommand{\omegawB}{^{\mathcal{B}}\boldsymbol{\omega}_\mathbf{w}}

\newcommand{\vp}{\boldsymbol{\xi}_p}
\newcommand{\vpB}{^{\mathcal{B}}\vp}
\newcommand{\vsB}{^{\mathcal{B}}\vs}
\newcommand{\vB}{{^{\mathcal{B}}\boldsymbol{\xi}}}
\newcommand{\velosB}{^{\mathcal{B}}\bar{\mathbf{v}}_{s}}
\newcommand{\velosE}{^{\mathcal{E}}\bar{\mathbf{v}}_{s}}
\newcommand{\velopB}{^{\mathcal{B}}\bar{\mathbf{v}}_{p}}

\newcommand{\vpE}{^{\mathcal{E}}\vp}
\newcommand{\vsE}{^{\mathcal{E}}\vs}
\newcommand{\vE}{{^{\mathcal{E}}\boldsymbol{\xi}}}
\newcommand{\vwE}{^{\mathcal{E}}\boldsymbol{\xi}_{\mathbf{w}}}
\newcommand{\vwB}{^{\mathcal{B}}\boldsymbol{\xi}_{\mathbf{w}}}
\newcommand{\Dt}{\Delta t}

\newcommand{\rbar}{\bar{\mathbf{r}}}
\newcommand{\vbar}{\bar{\mathbf{v}}}

\newcommand{\sigmarcs}{\sigma_{\text{RCS}}}
\newcommand{\ones}{\mathbf{1}}
\newcommand{\aT}{\mathbf{a}_{T}}
\newcommand{\aR}{\mathbf{a}_{R}}
\newcommand{\aTx}{\mathbf{a}_{T,x}}
\newcommand{\aTy}{\mathbf{a}_{T,y}}

\newcommand{\nTx}{\mathbf{n}_{T,x}}
\newcommand{\nTy}{\mathbf{n}_{T,y}}
\newcommand{\nRx}{\mathbf{n}_{R,x}}
\newcommand{\nRy}{\mathbf{n}_{R,y}}
\newcommand{\Rebrk}[1]{\mathbf{Re}\{ #1 \}}
\newcommand{\Imbrk}[1]{\mathbf{Im}\{ #1 \}}

\newcommand{\rank}{\operatorname{rank}}
\newcommand{\logdet}{\operatorname{log \, det}}
\newcommand{\B}{\mathbf{B}}

\newcommand{\coneone}{\mathbf{c}_{11}}
\newcommand{\conetwo}{\mathbf{c}_{12}}

\newcommand{\ctwo}{\mathbf{c}_2}
\newcommand{\conet}{\mathbf{c}_1^\top}
\newcommand{\ctwot}{\mathbf{c}_2^\top}
\newcommand{\coneonet}{\mathbf{c}_{11}^\top}
\newcommand{\conetwot}{\mathbf{c}_{12}^\top}
\newcommand{\Prmatperp}{\mathbf{P}_{\mathbf{r}}^\perp}
\newcommand{\Dmat}{\mathbf{D}}

\usepackage[cmex10]{amsmath} 

\newtheorem{rem}{Remark}

\begin{document}
\title{$SE(3)$-Based Trajectory Optimization and Target Tracking in UAV-Enabled ISAC Systems \\
  \thanks{This work was partly supported by the German Federal Ministry of Education and Research (BMBF) within the national initiative on 6G Communication Systems through the research hub \textit{6G-life} under Grant 16KISK002, within the project \textit{QUIET} under Grand 16KISQ093, and within the project \textit{QD-CamNetz} under Grand 16KISQ077. It was also supported by the Bavaria State Minister for Digital Affairs (StMD) within the project \textit{Next Generation AI Computing (GAIn)}.}
}

\author{%
  \IEEEauthorblockN{Dongxiao Xu${ }^*$, Xinyang Li${ }^*$, Vlad C. Andrei${ }^*$, Moritz Wiese${ }^*$, Ullrich J. Mönich${ }^*$, Holger Boche${ }^*$}
  \IEEEauthorblockA{\textit{*Chair of Theoretical Information Technology} \\
  \textit{Technical University of Munich}\\
  Munich, Germany \\
Emails: \{dongxiao.xu, xinyang.li, vlad.andrei, wiese, moenich, boche\}@tum.de}
}

\maketitle


\begin{abstract}
This paper presents a novel approach to enhance sensing capabilities in UAV-enabled MIMO-OFDM ISAC systems by leveraging UAV mobility as a mono-static radar. By integrating uniform planar arrays (UPAs) and modeling the UAV dynamics in \( SE(3) \), we address key challenges such as 3D space sensing and trajectory design. We propose a target tracking scheme using extended Kalman filtering (EKF) in \( SE(3) \), along with trajectory optimization based on the conditional Posterior Cramer-Rao bound (CPCRB). Numerical results demonstrate the effectiveness of the proposed trajectory design in enhancing performance of target tracking and physical parameter estimation in UAV-enabled MIMO-OFDM ISAC systems.
\end{abstract}
\section{Introduction}
Unmanned aerial vehicles (UAVs), serving as aerial base stations (BS), are expected to revolutionise the integrated sensing and communication (ISAC) systems with their high mobility and flexible deployment. Recent research has focused intensively on the joint beamforming and UAV trajectory optimisation \cite{meng2022uav}, \cite{lyu2022joint}, yet key limitations remain:

(i) Limitation of ULAs: Many prior works either do not specify antenna arrays \cite{jing2024isac}, \cite{wu2023interplay}, or assume uniform linear arrays (ULAs), mounted vertically \cite{lyu2022joint} or horizontally \cite{meng2022uav}. ULAs capture only one degree of freedom for angles of arrival (AoA) or departure (AoD), limiting their 3D sensing ability. In contrast, uniform planar arrays (UPA) inherently support 3D spatial relationships and can capture both polar and azimuthal angle information. (ii) Rigid-Body Dynamics: Many studies \cite{meng2022throughput}, \cite{jing2024isac} focus on target tracking and real-time trajectory design in the Euclidean space, often requiring linear approximations in the state evolution model. Such approximations can be effectively avoided by operating within the \textit{special Euclidean group}  \cite{mueller2019modern}, \cite{gallo2022so}, i.e., $SE(3)$. (iii) Trajectory Design: Trajectories, typically as sequences of future states and actions, are typically explored in dynamic programming (DP) or model predictive control (MPC). However, state-of-the-art methods \cite{jing2024isac} primarily focus on optimizing UAV positions while overlooking control design aspects, which are crucial for practical implementation. (iv) Optimisation w.r.t. Posterior Cramer-Rao Bound (PCRB): While PCRB is commonly used to quantify estimation accuracy, prior works focus on optimizing power and bandwidth \cite{dong2022sensing}, beamforming \cite{liu2020radar}, \cite{liu2021cramer}, or waveform design \cite{li2023optimal}, few works have considered trajectory optimization based on PCRB metrics, especially in the framework of UAV-enabled ISAC systems.

This work addresses the limitations above by proposing a unified framework for UAV-enabled MIMO-OFDM ISAC systems that incorporates a practical UPA configuration, models UAV dynamics in $SE(3)$ for accurate state evolution, and fully exploits UAV mobility in trajectory design to enhance target tracking and physical parameter estimation performance.

\section{System Model}
\subsection{Radar Signal Model}
Considering a MIMO-OFDM ISAC system, where the UAV is equipped with a half wavelength spacing UPA with $N_T$ and $N_R$ antennas. The steering vector of the UPA is expressed as
\begin{gather}
		\mathbf{a}\left(\theta, \phi\right)=\mathbf{a}_y\left(\theta, \phi\right)\otimes \mathbf{a}_x\left(\theta, \phi\right) \in\mathbb{C}^{N},\\
	\resizebox{0.9\hsize}{!}{$
		\mathbf{a}_x\left(\theta, \phi\right)=\frac{1}{\sqrt{N_x}}\big[1, -e^{j\pi\sin\theta\cos\phi}, \ldots, -e^{j\pi\left(N_x-1\right)\sin\theta\cos\phi}\big]^\top, 
		$} \\
	\resizebox{0.9\hsize}{!}{$
		\mathbf{a}_y\left(\theta, \phi\right)=\frac{1}{\sqrt{N_y}}\big[1, -e^{j\pi\sin\theta\sin\phi}, \ldots, -e^{j\pi\left(N_y-1\right)\sin\theta\sin\phi}\big]^\top, 
		$}
\end{gather}
where $\otimes$ denotes the Kronecker product, $\theta$ and $\phi$ the polar angle and azimuthal angle, respectively, and $N=N_xN_y$. The Tx and Rx array responses share the structure, i.e., $	\mathbf{a}_{T}\left(\theta, \phi\right)=\mathbf{a}_{T,y}\left(\theta, \phi\right)\otimes \mathbf{a}_{T,x}\left(\theta, \phi\right) \in\mathbb{C}^{N_T}$, and $\mathbf{a}_{R}\left(\theta, \phi\right)=\mathbf{a}_{R,y}\left(\theta, \phi\right)\otimes \mathbf{a}_{R,x}\left(\theta, \phi\right) \in\mathbb{C}^{N_R}$, where $N_T=N_{T,x}N_{T,y}$, $N_R=N_{R,x}N_{R,y}$. The UAV communicates with a single GU while performing sensing by collecting signals reflected from the GU, as a mono-static radar, thus, AoA approximates AoD.  Assuming a single line-of-sight (LOS) path, the channel frequency response at subcarrier $\ell$ and OFDM symbol $k$ is
\begin{equation}
  \mathbf{H}_{\ell,k}=b\cdot\omegatilde_{\ell,k}\cdot\mathbf{a}_{R}(\theta, \phi)\mathbf{a}_{T}(\theta, \phi)^{\Herm},
\end{equation}
with $\omegatilde_{\ell, k}=e^{-j 2 \pi \ell f_0 \tau} e^{j 2 \pi \mu k T_s}$, where $T_s=1/ f_0+T_g$ is the OFDM symbol duration, with $T_g$ the cyclic prefix duration, $f_0$ the subcarrier spacing; $\tau$ is the path delay; $\mu$ is the doppler shift due to mobilities; $\mathbf{a}_T\left(\theta,\phi\right) \in \mathbb{C}^{N_T}$ and $\mathbf{a}_R\left(\theta,\phi\right) \in \mathbb{C}^{N_R}$ denote the UPA response at $\operatorname{AoD}$ and $\operatorname{AoA}$$\big(\theta, \phi\big)$, respectively; $b \in \mathbb{C}$ is the LOS-channel gain, which comprises the amplitude attenuation and phase shift, i.e., $b=\left| b \right|\cdot e^{j\varphi}=a \cdot d^{-2} \cdot e^{j\varphi}$, where $a\triangleq\sqrt{\lambda_c^2 \sigma_{\mathrm{rcs}} / (4 \pi)^3}\in\mathbb{R}$, with $\sigmarcs$ the radar cross section (RCS), $\lambda_c$ the wavelength of the centre frequency $f_c$. Strictly speaking, the phase shift $\varphi$ depends on the path delay \cite{braun2014ofdm}, but for $d\gg\lambda_c$, it can be considered as a uniformly distributed random variable \cite{tse2005fundamentals}, i.e., $\varphi\sim\mathcal{U}(-\pi, \pi)$.

Suppose the UAV transmits a signal $\mathbf{x}_{\ell, k}$ at the $(\ell, k)$-th OFDM resource element (RE) of size $L \times K$. The reflected signal received at the UAV is $\mathbf{y}_{\ell, k}=\mathbf{H}_{\ell, k} \mathbf{x}_{\ell, k}+\mathbf{z}_{\ell, k}$, with $\mathbf{z}_{\ell, k}\in\mathbb{C}^{N_R}$ being the additive gaussian noise of zero mean and covariance matrix $\mathbf{C}_{\mathbf{z}_{\ell, k}}=\sigma_{z}^2 \mathbf{I}_{N_R}$, i.e., the noise on different receive atennas is assumed to be independent. By using the property of the kronecker product, we have $\mathbf{y}_{\ell, k}=\big(\mathbf{x}_{\ell, k}^{\top} \otimes \mathbf{I}_{N_R}\big) \mathbf{h}_{\ell, k}+\mathbf{z}_{\ell, k}$, where $\mathbf{h}_{\ell, k}\triangleq\mathbf{vec}\left(\mathbf{H}_{\ell, k}\right)= b\cdot\omegatilde_{\ell,k}\cdot \mathbf{a}_{T}^*(\theta, \phi) \otimes \mathbf{a}_{R}(\theta, \phi)$. Without loss of generality, the sensing symbols are allocated on $M$ OFDM REs, i.e., $\left\{\left(\ell_m, k_m\right)\right\}_{m=1}^M$. For simplicity, the subscript $m$ is used to represent $\left(\ell_m, k_m\right)$. All $M$ received symbols can be further stacked vertically, i.e., $\mathbf{y}\triangleq\left[\mathbf{y}_1^\top,\cdots,\mathbf{y}_M^\top\right]^\top =\left(\mathbf{X}\otimes\I_{N_R}\right)\cdot\mathbf{h}+\mathbf{z}$, with $\mathbf{X}\triangleq\operatorname{blkdiag}\left(\mathbf{x}_1^\top,\ldots,\mathbf{x}_M^\top\right)\in\mathbb{C}^{M\times MN_T}$, $\mathbf{h}\triangleq\left[\mathbf{h}_1^\top,\cdots,\mathbf{h}_M^\top\right]^\top\in\mathbb{C}^{MN_RN_T}$, and  $\mathbf{z}\triangleq\left[\mathbf{z}_1^\top,\cdots,\mathbf{z}_M^\top\right]^\top\in\mathbb{C}^{MN_R}$, assuming independent noise on different REs, i.e., $\mathbf{C}_{\mathbf{z}}=\sigma_{z}^2 \mathbf{I}_{MN_R}$. Besides, $\mathbf{h}$ has a compact form given as $\mathbf{h}=b\cdot\boldsymbol{\omegatilde}\otimes\mathbf{a}_{T}^*(\theta, \phi) \otimes \mathbf{a}_{R}(\theta, \phi)$, with $\boldsymbol{\omegatilde}\triangleq\left[\omegatilde_1,\cdots,\omegatilde_M\right]^\top\in\mathbb{C}^{M}$, and $\omegatilde_m=e^{-j 2 \pi \ell_m f_0 \tau} e^{j 2 \pi \mu k_m T_s}$. The channel dependence on the physical parameters is expressed as \(\mathbf{h}(\boldsymbol{\zeta})\), where \(\boldsymbol{\zeta} = [\tau, \phi, \theta, \mu]^\top\) as the collection of all physical parameters.

 \subsection{State Evolution Model}
Hereby, we model the spatial configuration of the GU relative to the UAV as the state variable, leveraing $SE(3)$ to represent rigid-body motions, comprising both rotation $\Rmat$ and translation $\rvec$. One element of it can be expressed as a \textit{homogeneous transformation matrix} \cite{mueller2019modern}, i.e., $\T\triangleq(\Rmat,\rvec)$ for brevity. In this work, we consider one stationary \textit{global frame} $\{w\}$ fixed arbitrarily to the ground in an open 3D space. The UAV is modeled as a \textit{rigid body} with a \textit{body frame} $\{s\}$ located at its rotation center; while the GU is treated as a point with a body frame $\{p\}$ at its center. We use the following notations: (i) \(\mathbf{T}_{w, s}^n\): the configuration of the UAV's body frame $\{s_n\}$ relative to the global frame $\{w\}$ at the \(n\)-th epoch, (ii) \(\mathbf{T}_{w, p}^n\): the configuration of the GU's body frame $\{p_n\}$ relative to the global frame $\{w\}$ the \(n\)-th epoch, and (iii) \(\mathbf{T}_{s, p}^n\): the configuration of the GU's body frame $\{p_n\}$ relative to the UAV's body frame $\{s_n\}$ at the \(n\)-th epoch, which is expressed as $\mathbf{T}_{s,p}^n=\mathbf{T}_{s,w}^{n}\mathbf{T}_{w,p}^{n}=\left( \mathbf{T}_{w,s}^{n}\right) ^{-1}\mathbf{T}_{w,p}^{n}$, with $n=1,\ldots, N_e$. The configurations $\mathbf{T}_{w, s}^n$ and $\mathbf{T}_{w, p}^n$ evolve according to the rigid-body motions of the UAV and the GU, respectively, i.e.,
\begin{equation*}
  \begin{aligned}
    \Tws^n = \Tws^{n-1}\cdot\Exp\big( {\vsB^n} \Dt\big) = \Exp\big( {\vsE^n} \Dt\big)\cdot\Tws^{n-1}, \\
    \Twp^n = \Twp^{n-1}\cdot\Exp\big( {\vpB^n} \Dt\big) = \Exp\big( {\vpE^n} \Dt\big)\cdot\Twp^{n-1}, \\
  \vpB^n= \begin{bmatrix}
    {^{\mathcal{B}}\boldsymbol{\nu}_p^{n,\top}}, \boldsymbol{0}^\top
  \end{bmatrix}^\top, \ \vsB^n=\begin{bmatrix}
	^{\mathcal{B}}\boldsymbol{\nu}_s^{n,\top} \ ^{\mathcal{B}}\boldsymbol{\omega}_s^{n,\top}
	\end{bmatrix}^\top\in\mathbb{R}^6,
  \end{aligned}
\end{equation*}
where, in general, ${^{\mathcal{E}}\mathbf{v}}$ denotes the tangent element at the identity element $\mathcal{E}$ of $SE(3)$, expressed in the \textit{spatial} or \textit{global} frame $\mathcal{E}$, which can be transformed from a local element ${^{\mathcal{B}}\mathbf{v}}$ in its \textit{body} frame $\mathcal{B}$, via the \textit{adjoint action} \cite{gallo2022so}. In this work, the terminology of $w$ and $\mathcal{E}$ are used interchangeably. $\Exp(\cdot)$ indicates the \textit{capitalized exponential map} \cite{sola2021micro}, which maps a vector element $\mathbf{v}$, as an isomorphism to the element $\mathbf{v}^{\wedge}$ of the Lie algebra $se(3)$, to an element of $SE(3)$, where $(\cdot)^\wedge$ denotes the \textit{hat operator} \cite{gallo2022so}. The Lie algebra velocity $\mathbf{v}^\wedge$ of $SE(3)$ is known as \textit{twist} $\boldsymbol{\xi}^\wedge$, which represents the motion velocity in terms of both \textit{linear} $\boldsymbol{\nu}$ and \textit{angular velocity} $\boldsymbol{\omega}$. Naturally, we have the \textit{spatial} and \textit{body twist} as $\vE$ and $\vB$, respectively. 

In our case, the body twist of the UAV at the $n-1$-th epoch $\vsB^n$, consists of the linear and angular velocities, i.e., $^{\mathcal{B}}\boldsymbol{\nu}_s^n, {^{\mathcal{B}}\boldsymbol{\omega}_s^n}\in\mathbb{R}^3$, corresponding to a tangent vector at $\mathbf{T}_{w,s}^{n-1}$; while that of the GU at the identical epoch $\vpB^n$, comprises the linear velocity, i.e., $^{\mathcal{B}}\boldsymbol{\nu}_p^n\in\mathbb{R}^3$, and zero angular velocity, as the GU is treated as a point, corresponding to a tangent vector at $\mathbf{T}_{w,p}^{n-1}$,. In contrast, $\vsE^n$ and $\vpE^n$ are their counterparts in the spatial frame, i.e., ${\vsE^n}, {\vpE^n}\in se(3)$. In practice, the \textit{control signal} $\vsB^n$, as a body twist, is corrputed by a time-independent additive Gaussian noise \cite{sola2021micro}, $\R^6\ni{\vwB}=\begin{bmatrix}
	\nuwB^\top \ \omegawB^\top
\end{bmatrix}^\top\sim\mathcal{N}\left(\mathbf{0},\mathbf{\Xi}_{\mathbf{w}}\right)$, with $\mathbf{\Xi}_{\mathbf{w}}=\diag\big(\sigma_{\nu_{\xi,x}}^2,\sigma_{\nu_{\xi,y}}^2,\sigma_{\nu_{\xi,z}}^2,\sigma_{\omega_{\xi,x}}^2,\sigma_{\omega_{\xi,y}}^2,\sigma_{\omega_{\xi, z}}^2\big)\in\R^{6\times 6}$. Therefore, the relative spatial configurations evolve as
\begin{equation}
	\resizebox{0.892\hsize}{!}{$
	\mathbf{T}_{s,p}^n = \Exp\left( -\left(\vsB^n+{\vwB}\right) \Dt\right)\cdot\Tsp^{n-1}\cdot\Exp\left( \vpB^n \Dt\right). \label{eq.1.42}
	$}
\end{equation}
\subsection{Measurement Model}
Similarly, the measurement model can be interpreted from UAV's perspective relative to the GU. Specifically, the reflected signal received at the UAV is reformulated as
\begin{gather} 
	\mathbf{y}_{s,p}^n=\left(\mathbf{X}^n\otimes\I_{N_R}\right)\cdot\mathbf{h}_{s,p}^n+\mathbf{z}_{s,p}^n, \label{eq.1.17} \\
	\mathbf{h}_{s,p}^n=b_{s,p}^n\cdot\boldsymbol{\omegatilde}_{s,p}^n\otimes\mathbf{a}_{T}(\theta_{s,p}^n, \phi_{s,p}^n) \otimes \mathbf{a}_{R}(\theta_{s,p}^n, \phi_{s,p}^n), \\
	\boldsymbol{\omegatilde}_{s,p}^n\triangleq\left[\omegatilde_{1,s,p}^n,\cdots,\omegatilde_{M,s,p}^n\right]^\top, \\
	\omegatilde_{m,s,p}^n=e^{-j 2 \pi \ell_m f_0 \tau_{s,p}^n} e^{j 2 \pi \mu_{s,p}^n k_m T_s},
\end{gather}
with parameters extracted from $\mathbf{T}_{s,p}^n\triangleq\big(\Rmat_{s,p}^n,\rvec_{s,p}^n\big)$, i.e.,
\begin{align}
	\theta_{s,p}^n & = \arccos(z_{s,p}^n / \rho_{s,p}^n ), \label{eq.1.37}\\
	\phi_{s,p}^n &= \operatorname{atan2} \left( y_{s,p}^n, x_{s,p}^n\right), \label{eq.1.38} \\
	\tau_{s,p}^n &= 2\cdot\rho_{s,p}^n / c, \label{eq.1.36} \\
	b_{s,p}^n &= a\cdot e^{j\varphi_{s,p}^n}/(\rho_{s,p}^{n})^2, \label{eq.1.43} \\
  v_{\text{radial}}^n &= \left\langle\mathbf{R}_{s,p}^n{^{\mathcal{B}} \mathbf{v}_p^n}-{^{\mathcal{B}} \mathbf{v}_s^n}, \mathbf{r}_{s,p}^n\right\rangle / \rho_{s,p}^n, \label{eq.1.44} \\
  \mu_{s,p}^n&=2\cdot v_{\text{radial}}^n \cdot f_c / c, \label{eq.1.33}
\end{align}
where $\mathbf{r}_{s,p}^n=\big[ x_{s,p}^n, y_{s,p}^n, z_{s,p}^n\big]^\top$, $\rho_{s,p}^n = \|\mathbf{r}_{s,p}^n\|$, the scalar $c$ is the speed of light, and the phase shift is assumed to be wide-sense-stationary (WSS), i.e., $\varphi_{s,p}^n=\varphi\sim\mathcal{U}(-\pi, \pi)$. $^{\mathcal{B}} \mathbf{v}_s^n\in\R^3$ indicates the local velocity of the UPA installed on the UAV in $\{s_n\}$, which depends on the control signal $\vsB^{n+1}$ and the midpoint coordinates of the UPA, i.e., $\mathbf{s} \in \mathbb{R}^3$, relative to the origin of $\{s_n\}$, in the given form $\big({ }^{\mathcal{B}} \boldsymbol{\xi}_s^{n+1}\big)^{\wedge} \overline{\mathbf{s}}=\big[{ }^{\mathcal{B}} \mathbf{v}_s^{n, \top}, 0\big]^{\top}$, where $\overline{\mathbf{s}}=\big[\mathbf{s}^{\top}, 1\big]^{\top} \in \mathbb{R}^4$ is the \textit{homogeneous coordinates} of $\mathbf{s}$. Similarly, $^{\mathcal{B}} \mathbf{v}_p^n\in\R^3$ indicates the local velocity of the GU in $\{p_n\}$, given by $\big({ }^{\mathcal{B}} \boldsymbol{\xi}_p^{n+1}\big)^{\wedge} \overline{\mathbf{p}}=\big[{ }^{\mathcal{B}} \mathbf{v}_p^{n, \top}, 0\big]^{\top}$, where $\overline{\mathbf{p}}=\left[\mathbf{0}^{\top}, 1\right]^{\top} \in \mathbb{R}^4$. Equations \eqref{eq.1.44} and \eqref{eq.1.33} show that the Doppler frequency shift is influenced by the control signal $\vsB^{n+1}$. Besides, (\ref{eq.1.37}) to (\ref{eq.1.33}) highlight the dependence of physical parameters on the relative configuration, thus, we express $\boldsymbol{\zeta}_{s,p}^n$ as $\boldsymbol{\zeta}_{s,p}^n\big(\Tsp^n, {\vsB^{n+1}}\big)$.
\begin{proof}
  Eq. \eqref{eq.1.37} to \eqref{eq.1.43} are straightforward. For the proof of \eqref{eq.1.44}, please refer to the Appendix \ref{sub.1.4}.
\end{proof}

\section{The Proposed Approach for GU Tracking}
\subsection{Extended Kalman Filtering in $SE(3)$}
In this subsection, an Extended Kalman Filtering (EKF) scheme in $SE(3)$ is proposed for GU tracking. The state evolution model is given by (\ref{eq.1.42}), where $\mathbf{f}_n: SE(3)\times \R^6 \times \R^6 \to SE(3)$, while the measurement model by (\ref{eq.1.17}), where $\mathbf{g}_n:SE(3)\times \R^6 \to\mathbb{C}^{MN_R}$, summarised as
\begin{equation}
  \left\{\begin{aligned}
    \Tsp^n &= \mathbf{f}_n\left(\Tsp^{n-1},{\vs^{n}}, {\boldsymbol{\xi}_{\mathbf{w}}} \right)\oplus\mathbf{w}, \\
    \mathbf{y}_{s,p}^n &= \mathbf{g}_n\left(\Tsp^n,{\vs^{n+1}}\right) + \mathbf{z},
	\end{aligned}\right.
\end{equation}
where $\vs^n$ denotes $\vsB^n$ or $\vsE^n$, and the additive process noise is defined with the \textit{right-plus} operator $\oplus$, denoted as $\R^6\ni\mathbf{w}\sim\mathcal{N}\left(\mathbf{0},\mathbf{C}_{\mathbf{w}}\right)$ with $\mathbf{C}_{\mathbf{w}}=\diag\big(\sigma_{\nu_{x}}^2,\sigma_{\nu_{y}}^2,\sigma_{\nu_{z}}^2,\sigma_{\omega_{x}}^2,\sigma_{\omega_{y}}^2,\sigma_{\omega_{z}}^2\big)\in\R^{6\times 6}$. The dependence of $\ysp^n$ on $\Tsp^n$ is composed as $\ysp^n=\mathbf{g}_n\big(\hsp^n\big(\zetasp^n\big(\Tsp^n,\cdot\big)\big)\big)$. The EKF procedure for target tracking and parameter estimation follows, with the superscript $-$ denoting \textit{a priori} state estimates, and the absence indicating \textit{a posteriori} counterparts, in line with the notation in \cite{sarkka2023bayesian}.
\subsubsection{Prediction step}
At each epoch, we apply EKF prediction with equations: (i) $\Tsphat^{n,-}=\mathbf{f}_n(\Tsphat^{n-1},{\vs^n)}$. (ii) $\mathbf{P}_n^-=\F_{n-1}\mathbf{P}_{n-1}\F_{n-1}^\top+\mathbf{G}_{n-1}\mathbf{\Xi}_{\mathbf{w}}\mathbf{G}_{n-1}^\top+\mathbf{C}_{\mathbf{w}}$, with the Jacobians computed from the state evolution model, i.e.,
\begin{align}
  \F_{n-1}&\triangleq \left.\mathbf{J}_{\Tsp^{n-1}}^{\Tsp^n}\right|_{\mathbf{w}=\zero} =\mathbf{Ad}_{\Exp\left(-\vpB^n\Dt\right)}, \label{eq.1.68} \\
	\mathbf{G}_{n-1} &\triangleq\left. \mathbf{J}_{\vsB^{n}}^{\Tsp^n}\right|_{\Tsp^{n-1}=\Tsphat^{n-1},\vsB^n=\vsB^n, \boldsymbol{\xi}_{\mathbf{w}}=\zero, \mathbf{w}=\zero} \nonumber \\
	                 &=-\mathbf{Ad}_{\Tsphat^n}^{-1}\cdot\mathbf{J}_r(\vsB^n\Dt)\cdot\Dt, \\
    \mathbf{J}_{\Tsp^{n-1}}^{\Tsp^n}&=\mathbf{Ad}_{\Exp\left(-\vpB^n\Dt\right)\oplus\mathbf{w}}, \\
		\mathbf{J}_{\vwB}^{\Tsp^n}&=-\mathbf{Ad}_{\Tsp^n}^{-1}\cdot\mathbf{J}_r(\left(\vsB^n+{\vwB}\right)\Dt)\cdot\Dt.
\end{align}

Apart from the target tracking, the physical parameters $\boldsymbol{\zeta}_{s,p}^n$ can also be estimated in an intermediate step and subsequently utilized for beamforming design \cite{liu2020radar}. The conditional distribution of $\Tsp^n$ is characterised as $p(\Tsp^n\vert \ysp^{1:n-1})\sim\mathcal{N}\big(\Tsphat^{n,-}, \mathbf{P}_n^-\big)$, cf. Theorem 4.2 of \cite{sarkka2023bayesian}. Given the dependence of $\zetasp^n$ on $\Tsp^n$, we have the following distribution $p(\zetasp^n\vert \ysp^{1:n-1})\sim\mathcal{N}\big(\zetasphat^{n,-}, \mathbf{V}_n^-\big)$ with $\zetasphat^{n,-} = \zetasp\big(\Tsphat^{n,-}\big)$, $\mathbf{V}_n^- = \mathbf{\Psi}_n\mathbf{P}_n^-\mathbf{\Psi}_n^\top$, cf. Lemma A.1 of \cite{sarkka2023bayesian}, where the Jacobian $\mathbf{\Psi}_n$ is used to propagate uncertainty of $\Tsphat^{n,-}$ to $\zetasphat^{n,-}$, where 
\begin{equation} \label{eq.1.72}
	\mathbf{\Psi}_n \triangleq\left.\mathbf{J}_{\Tsp^n}^{\boldsymbol{\zeta}_{s,p}^n}\right|_{\Tsp^{n}=\Tsphat^{n,-},\vsB^{n+1}=\vsB^{n+1}}.
\end{equation}
\subsubsection{Correction step}
For brevity, the subscript $s,p$ and superscript $n$ in the sequel are omitted. Considering (\ref{eq.1.17}), we obtain $\J_{\mathbf{h}}^{\mathbf{y}} = \mathbf{X}\otimes\I_{N_R}\in\mathbb{C}^{MN_R\times MN_RN_T}$. By leveraging basic Jacobian blocks \cite{gallo2022so} in $SE(3)$, $\J_{\boldsymbol{\zeta}}^{\mathbf{h}}\in\mathbb{C}^{MN_RN_T\times 4}$ reads as
\begin{equation} \label{eq.1.59}
	\begin{aligned}
		&\J_{\boldsymbol{\zeta}}^{\mathbf{h}} = \begin{bmatrix}
			\J_{\tau}^{\mathbf{h}} & \J_{\phi}^{\mathbf{h}} & \J_{\theta}^{\mathbf{h}} & \J_{\mu}^{\mathbf{h}}
		\end{bmatrix} = j\pi b \cdot \big(\boldsymbol{\omegatilde} \otimes\aT^* \otimes \aR\big) \bullet \\
    & \begin{bmatrix}
    -2 f_0 \cdot \boldsymbol{\tilde{\ell}} & \sin\theta \cdot \big(\mathbf{N}_1 \mathbf{f}_{\phi}\big) & \cos\theta \cdot \big(\mathbf{N}_2 \mathbf{f}_{\phi}\big) & 2 T_s \cdot \mathbf{k}
    \end{bmatrix},
	\end{aligned}
\end{equation} 
where $\bullet$ indicates the face-splitting product. Components, e.g., $\mathbf{k}$, $\mathbf{N}_1$, depend on the RE and UPA configurations.
\begin{proof}
	 Please refer to Appendix \ref{sub.1.1}.
\end{proof}
The Jacobian $\J_{\boldsymbol{\zeta}}^{\mathbf{y}}$ is constructed via the chain rule, i.e., $\J_{\boldsymbol{\zeta}}^{\mathbf{y}} = \J_{\mathbf{h}}^{\mathbf{y}}\cdot \J_{\boldsymbol{\zeta}}^{\mathbf{h}}\in\mathbb{C}^{MN_R\times 4}$, in the complex domain. In order to enable the correction step in the Lie group setting, the measurement $\mathbf{y}=\Rebrk{\mathbf{y}} + j\cdot \Imbrk{\mathbf{y}}\in\mathbb{C}^{MN_R}$ is expressed in its argumented form $\underline{\mathbf{y}}=\big[\Rebrk{\mathbf{y}}^\top \ \Imbrk{\mathbf{y}}^\top\big]^\top\in\mathbb{R}^{2MN_R}$, with a similar expression for $\underline{\mathbf{z}}\in\mathbb{R}^{2MN_R}$. Thus, the derivative of $\underline{\mathbf{y}}$ w.r.t. $\boldsymbol{\zeta}$ reads as
$\J_{\boldsymbol{\zeta}}^{\underline{\mathbf{y}}}=\big[\Rebrk{\J_{\boldsymbol{\zeta}}^{\mathbf{y}}}^\top \ \Imbrk{\J_{\boldsymbol{\zeta}}^{\mathbf{y}}}^\top\big]^\top\in\mathbb{R}^{2MN_R\times 4}$. Besides, the structure of $\J_\T^{\boldsymbol{\zeta}}\in\R^{4\times 6}$ in \eqref{eq.1.72} is summarised as
\vspace{-10pt}
\begin{align}
  \J_\T^{\boldsymbol{\zeta}}&=\big[
		\J_\T^{\tau,\top} \ \J_\T^{\phi,\top} \ \J_\T^{\theta,\top} \ \J_\T^{\mu,\top}
  \big]^\top, \label{eq.1.60}\\
	\J_{\T}^\tau &= \frac{2}{c\rnorm}\begin{bmatrix}
		\rvec^\top\mathbf{R} & \mathbf{0}_{1\times 3} \\
	\end{bmatrix}, \\
	\J_{\T}^{\phi}&=\frac{1}{x^2+y^2}\begin{bmatrix}
		x\cdot\etwo^\top\mathbf{R}-y\cdot\eone^\top\mathbf{R} & \zero_{1\times 3}
	\end{bmatrix}, \\
	\J_\T^\theta &=-\frac{1}{\sqrt{\rnorm^2-z^2}}\begin{bmatrix}
		\ethree^\top\mathbf{R}-\frac{z\cdot\mathbf{r}^\top \mathbf{R} }{\rnorm^2} & \mathbf{0}_{1\times 3}
	\end{bmatrix}, \\
	\mathbf{J}^{\mu}_{\mathbf{T}}&=\frac{2f_c}{c\rnorm}\left(
	\begin{bmatrix}
		{^{\mathcal{B}}{\mathbf{v}}_{p}^\top}-{^{\mathcal{B}}{\mathbf{v}}_{s}^\top}\mathbf{R}
		& -\rvec^\top\mathbf{R}\left[{^{\mathcal{B}}{\mathbf{v}}_{p}}\right]_{\times}	\end{bmatrix} \right. \nonumber\\
	&\quad \left. -\rnorm^{-2}\cdot\langle \mathbf{R}{^{\mathcal{B}}{\mathbf{v}}_{p}} - {^{\mathcal{B}}{\mathbf{v}}_{s}}, \rvec\rangle\cdot\begin{bmatrix}
		\rvec^\top\mathbf{R} & \zero_{1\times 3}
	\end{bmatrix}
	\right).
\end{align}
\begin{proof}
Please refer to Appendix \ref{sub.1.0}.
\end{proof}
By considering the chain rule $\J_{\T}^{\underline{\mathbf{y}}}=\J_{\boldsymbol{\zeta}}^{\underline{\mathbf{y}}}\cdot\J_{\T}^{\boldsymbol{\zeta}}\in\R^{2MN_R\times 6}$, the Jacobian $\mathbf{H}_n\in\R^{2MN_R\times 6}$ is obtained as
\begin{equation} \label{eq.1.71}
	\mathbf{H}_n\triangleq\left.\mathbf{J}_{\Tsp^n}^{\underline{\mathbf{y}}_{s,p}^n}\right|_{\Tsp^{n}=\Tsphat^{n,-},\vsB^{n+1}=\vsB^{n+1}},
\end{equation}
which leads to correction equations: (i) $\mathbf{S}_n=\mathbf{H}_n\mathbf{P}_n^{-}\mathbf{H}_n^\top+\mathbf{C}_{\underline{\mathbf{z}}}$. (ii) $\mathbf{K}_n=\mathbf{P}_n^{-}\mathbf{H}_n^\top \mathbf{S}_n^{-1}$. (iii) $\Tsphat^{n} =\Tsphat^{n,-}\oplus\mathbf{K}_n\big(\underline{\mathbf{y}}_{s,p}^n-\underline{\mathbf{g}}_n\big(\Tsphat^{n,-}\big)\big)$. (iv) $\mathbf{P}_n = \mathbf{P}_n^{-}-\mathbf{K}_n\mathbf{S}_n\mathbf{K}_n^\top$.

\subsection{Evaluation of the conditional PCRB}
The PCRB, defined as the inverse of the Fisher information matrix (FIM) for a random vector-valued parameter \cite{vantrees2013detection}, retains its recursive structure when extended to Lie groups \cite{chahbazian2024recursive}, given as $\mathcal{I}\left(X_{n}\right)=\mathbf{D}_{n-1}^{22}-\mathbf{D}_{n-1}^{21}\left(\mathcal{I}\left(X_{n-1}\right)+\mathbf{D}_{n-1}^{11}\right)^{-1} \mathbf{D}_{n-1}^{12}$, where $X_n\in\mathcal{G}$ denotes the state variable defined on a Lie group $\mathcal{G}$. See \cite{chahbazian2024recursive} for more details about submatrices, e.g., $\mathbf{D}_{n-1}^{11}$. For brevity, the following notation is adopted $\mathbf{f}_n\big(\Tsp^{n-1} \big)\triangleq\mathbf{f}_n\big(\Tsp^{n-1},{\vs^{n}}, \zero \big)$, leading to an approximation via the Baker–Campbell–Hausdorff (BCH) formula \cite{hall2013lie}, i.e., $\Tsp^n\approx\mathbf{f}_n\big(\Tsp^{n-1}\big)\Exp\big(-\mathbf{A d}_{\T_{w,p}^{n-1}}^{-1}{\vwE}\Dt+\mathbf{w}\big)$. Assuming a concentrated Gaussian distribution \cite{chahbazian2024recursive}, the logarithm of the parametrised conditional PDFs are given as $\log p\big(\Tsp^{n}\vert \Tsp^{n-1}\big)=c_1+\frac{1}{2}\left\|\Log \left(\mathbf{f}_n(\Tsp^{n-1})^{-1}\Tsp^n\right)\right\|_{\mathbf{C}_{\mathbf{w}}^\prime}^2$, where $\mathbf{C}_{\mathbf{w}}^\prime\approx\Dt^2\cdot\mathbf{A d}_{\T_{w,p}^{n-1}}^{-1}\mathbf{\Xi}_{\mathbf{w}}\mathbf{A d}_{\T_{w,p}^{n-1}}^{-1,\top}+\mathbf{C}_{\mathbf{w}}$, and $c_1$ as a constant. Applying the generic expression of $\mathbf{D}_{n-1}^{11}$ leads to $	\mathbf{D}_{n-1}^{11}=\mathbf{F}_{n-1}^\top\mathbf{C}_{\mathbf{w}}^{\prime,-1}\mathbf{F}_{n-1}$, where the Jacobian $\mathbf{F}_{n-1}$ is identical to (\ref{eq.1.68}). Similarly, we obtain other components $\mathbf{D}_{n-1}^{12}=\mathbf{F}_{n-1}^\top\mathbf{C}_{\mathbf{w}}^{\prime,-1} = \mathbf{D}_{n-1}^{21,\top}$, and $\mathbf{D}_{n-1}^{22}=\mathbf{C}_{\mathbf{w}}^{\prime,-1}+\mathbf{H}_n^\top\mathbf{C}_{\underline{\mathbf{z}}}^{-1}\mathbf{H}_n$. By utilising the Woodbury matrix identity \cite{higham2002accuracy}, we have
\begin{equation}
    \resizebox{.892\hsize}{!}{$
      \mathcal{I}(\Tsp^n)=\mathbf{H}_n^\top\mathbf{C}_{\underline{\mathbf{z}}}^{-1}\mathbf{H}_n+\big(\mathbf{C}_{\mathbf{w}}^{\prime}+\mathbf{F}_{n-1}\mathcal{I}\big(\Tsp^{n-1}\big)^{-1}\mathbf{F}_{n-1}^\top\big)^{-1},
  $}
\end{equation}
which has the same form as the conditional PCRB (CPCRB) proposed in \cite{zuo2011conditional}, i.e., $\mathcal{I}(\Tsp^n \vert \mathbf{y}_{s,p}^{1: n-1})$, yielding $\mathrm{MSE}(\hat{\T}_{s,p}^n\vert \mathbf{y}_{s,p}^{1:n-1})\geq\mathcal{I}\left(\Tsp^{n}\right)^{-1}\triangleq\mathrm{CPCRB}\left(\Tsp^{n}\right)$. Besides, $\mathrm{CPCRB}(\boldsymbol{\zeta}_n)$ can be evaluated via reparameterisation \cite{kay1993fundamentals}, i.e., $\mathrm{CPCRB}(\boldsymbol{\zeta}_{s,p}^n)=\mathbf{\Psi}_n\mathcal{I}\left(\Tsp^{n}\right)^{-1}\mathbf{\Psi}_n^\top\succ\zero$.

\section{Problem Formulation for Control Design}

This work focuses on three tasks: (i) Target tracking, assessed by the estimation of relative spatial configuration $\mathbf{T}_{s, p}^n$. (ii) Physical parameter estimation, relying on the estimation quality of $\mathbf{T}_{s, p}^n$. (iii) Control signal design, leveraging UAV mobility to improve tasks (i) and (ii), whose performance can be evaluated via $\operatorname{CPCRB}\left(\mathbf{T}_{s, p}^n\right)$ and $\operatorname{CPCRB}\left(\boldsymbol{\zeta}_{s, p}^n\right)$, respectively, with optimization criteria \cite{li2007range}, e.g., Trace-Opt. Given the dependence between (i) and (ii) on $\mathbf{T}_{s, p}^n$, control signal is optimized using Logdet-Opt w.r.t. CPCRB $\left(\mathbf{T}_{s,p}^n\right)$. The optimization problem is formulated as
\vspace{0pt}
\begin{align}
  \mathrm{P_{1}}: & \quad \min_{\vs^{n+1}} \logdet\big(\mathrm{CPCRB}(\T_n)\big) \label{eq.1.73} \\
  \mathrm{s.t.} & \quad \|\nuvec_s^{n+1}\|\leq V_{\ell}, \quad \|\nuvec_s^{n+1} - \nuvec_s^{n}\|\leq A_{\ell},\nonumber \\
                & \quad \|\omegavec_s^{n+1}\|\leq V_{a}, \quad \|\omegavec_s^{n+1} - \omegavec_s^{n}\|\leq A_{a}, \label{eq.1.86} \\
                & \quad \|{\velosE^{n}}-{\velosE^{n-1}}\|\leq V, \ \nu_{s,3}^{n+1}=\omega_{s,1}^{n+1}=\omega_{s,2}^{n+1}=0, \nonumber
\end{align}
where $\T_n\triangleq\Tsp^n$ for brevity. $V_\ell$ limits the UAV's maximum linear velocity in the horizontal plane at the $n$-th epoch, while $A_\ell$ bounds the change in linear velocity between consecutive epochs, limiting acceleration. $V_a$ and $A_a$ similarly constrain the UAV's angular velocity and acceleration about its vertical axis. $V$ restricts deviations in the global velocity of the UPA's midpoint, ensuring smooth trajectories, control stability, and reliable sensing. Lastly, the equality constraints fix the altitude, prevent tilting, and limit rotation to the vertical axis. The prior information FIM \cite{zhang2020power}, independent of the control signal, is denoted as $\mathbf{E}\triangleq\left(\mathbf{C}_{\mathbf{w}}^{\prime}+\mathbf{F}_{n-1}\mathcal{I}(\T_{n-1})^{-1}\mathbf{F}_{n-1}^\top\right)^{-1}\succ\zero$. Since only \(\boldsymbol{\Psi}_n\) in \eqref{eq.1.72} as part of \(\mathbf{H}_n\) depends on $\vs^{n+1}$, it can be separated via $\mathbf{A}\triangleq\J_{\boldsymbol{\zeta}}^{\underline{\mathbf{y}},\top}\J_{\boldsymbol{\zeta}}^{\underline{\mathbf{y}}}\succ\zero$, enabling the reformulation of \eqref{eq.1.73} into: $\min_{\vs^{n+1}} \logdet\big(\big(\mathbf{E}+\sigma_z^{-2}\mathbf{\Psi}_n^\top\mathbf{A}\mathbf{\Psi}_n\big)^{-1}\big)=\min_{\vs^{n+1}} -\logdet\big(\tilde{\mathbf{A}}^{-1}+\mathbf{\Psi}_n\mathbf{E}^{-1}\mathbf{\Psi}_n^\top\big)$. where $\tilde{\mathbf{A}}\triangleq\sigma_z^{-2}\mathbf{A}$. By isolating the component of $\mathbf{\Psi}_n$ that depends on $\vs^{n+1}$, problem $(\mathrm{P}_1)$ is reformulated as
\begin{equation}
  \mathrm{P_{2}}: \quad \min_{\vs^{n+1}} -\logdet\big(\tilde{\mathbf{A}}^{-1}+\mathbf{D}_0+\mathbf{D}\big) \quad \mathrm{s.t.} \quad (\ref{eq.1.86}),
\end{equation}
where $\mathbf{D}_0$ are independent of $\vs^{n+1}$, while $\mathbf{D} = \mathbf{e}_4\vs^{n+1,\top}\mathbf{K}^\top+\mathbf{K}\vs^{n+1}\mathbf{e}_4^\top+\vs^{n+1,\top}\mathbf{\tilde{\mathbf{P}}}\vs^{n+1}\mathbf{e}_4\mathbf{e}_4^\top$, $\Kmat \triangleq \Etilde^\top\Mmat\Smat$, $\Ptilde \triangleq \Smat^\top\Mmat^\top\mathbf{E}_{11}\Mmat\Smat\succcurlyeq\zero$, $\Smat \triangleq \begin{bmatrix} -\left[\mathbf{s}\right]_\times & \I \\ \end{bmatrix}$, with $\Etilde \triangleq \begin{bmatrix} \mathbf{E}_{11}\mathbf{B}^\top & \mathbf{E}_{11}\mathbf{c}_{11} + \mathbf{E}_{12}\mathbf{c}_{2} \\ \end{bmatrix}$, $\Mmat \triangleq -\frac{2f_c}{c\|\rvec\|}\cdot \mathbf{R}^\top\Prmatperp$. The matrices $\mathbf{E}_{11}$ and $\mathbf{E}_{12}$ reads directly from $\mathbf{E}^{-1}$, while $\mathbf{B}$, $\coneone$, and $\ctwo$ are the components from $\mathbf{\Psi}_n$. The reformulation is provided in Appendix \ref{sub.1.2}. As $\mathbf{D}$ is convex w.r.t. $\vs^{n+1}$, given that $\bar{\mathbf{P}}\succcurlyeq\zero$, but the outer $\logdet(\cdot)$ is concave over $\mathbb{S}_{++}^4$, the overall function becomes non-convex. Though $\mathbf{D}$ is a $4\times 4$ matrix, its sparsity (non-zero entries only in the last row and column) allows $(\mathrm{P}_2)$ to be reformulated as
\begin{equation}
  \mathrm{P_{2}^\prime}: \quad \max_{\vs^{n+1}} \ \vs^{n+1,\top}\mathbf{\bar{\mathbf{P}}}\vs^{n+1}+2\mathbf{c}^\top\boldsymbol{\xi}_s^{n+1}, \quad \mathrm{s.t.} \quad (\ref{eq.1.86}),
\end{equation}
with $\mathbf{c}\triangleq \Smat^\top\Mmat^\top\left(\mathbf{E}_{11}\coneone+\mathbf{E}_{12}\ctwo-\mathbf{E}_{11}\Bmat^\top\Abar^{-1}\mathbf{a}\right)$, and $\bar{\mathbf{P}}=\Smat^\top\Mmat^\top\mathbf{E}_{11}^{\frac{1}{2}}\big(\mathbf{I}-\mathbf{E}_{11}^{\frac{1}{2}}\Bmat^\top\bar{\mathbf{A}}^{-1}\Bmat\mathbf{E}_{11}^{\frac{1}{2}}\big)\mathbf{E}_{11}^{\frac{1}{2}}\Mmat\Smat$, where $\bar{\mathbf{A}}$ reads directly from $\tilde{\mathbf{A}}^{-1}+\mathbf{D}_0$. The reformulation is provided in Appendix \ref{sub.1.3}. Problem $(\mathrm{P}_2^\prime)$ can be further expressed as quadratic constrained quadratic programming (QCQP), i.e.,
\vspace{0pt}
\begin{align}
    \mathrm{P_{3}}:  \quad & \max_{\vs\in\R^6} \ \vs^{\top}\mathbf{Q}_0\vs+2\mathbf{c}^\top\vs \label{eq.1.85} \\
    \mathrm{s.t.}  \quad & \vs^{\top}\mathbf{Q}_i\vs\leq V_{i}, \ i = 1,2, \quad \vs^{\top}\mathbf{Q}_i\vs = 0, \ i = 4,5,6,  \nonumber\\ 
    \quad & \vs^{\top}\mathbf{Q}_i\vs-2\mathbf{b}_i^{\top}\vs+\|\mathbf{b}_i\|^2\leq A_{i}, \quad i = 1,2, \nonumber \\
    \quad & \vs^{\top}\mathbf{Q}_3\vs-2\mathbf{b}_3^{\top}\Rmat_2^\top\Rmat_1\Smat\vs+\|\mathbf{b}_3\|^2\leq V^2, \nonumber \\
    \mathrm{with} \quad & \mathbf{b}_i = \mathbf{Q}_i\vs^n, \ \mathbf{b}_3 = \Smat\vs^n, \qquad i = 1,2, \nonumber \\
    \quad & \Rmat_2 = \Rmat_{w,s}^{n-1}, \ \Rmat_1 = \Rmat_{w,s}^{n}, \ \mathbf{Q}_i=\mathbf{e}_i\mathbf{e}_i^\top, \quad i = 4,5,6, \nonumber \\
    \quad &  \mathbf{Q}_0=\bar{\mathbf{P}}, \ \mathbf{Q}_1 = \begin{bmatrix} \mathbf{I}_2 &  \\ & \zero \end{bmatrix}, \ \mathbf{Q}_2 = \mathbf{e}_6\mathbf{e}_6^\top, \ \mathbf{Q}_3 = \Smat^\top\Smat, \nonumber
\end{align}
where $V_1 = V_{\ell}^2, \ V_2 = V_{a}^2, \ A_1 = A_{\ell}^2, \ A_2 = A_{a}^2$, and $\vs\triangleq\vs^{n+1}$ for brevity. The objective function in (\ref{eq.1.85}) is concave, if and only if $\mathbf{I}\preccurlyeq\mathbf{E}_{11}^{\frac{1}{2}}\Bmat^\top\bar{\mathbf{A}}^{-1}\Bmat\mathbf{E}_{11}^{\frac{1}{2}}$, requiring all eigenvalues $\lambda_i\geq1$, which, however, does not always hold true, i.e., $\mathbf{Q}_0$ is solely symmetric, i.e., $\mathbf{Q}_0\in\mathbb{S}^6$, and $ \mathbf{Q}_i\succcurlyeq\zero, i=1,\ldots,6$. Therefore, despite the reduced dimensionality, $(\mathrm{P}_3)$ remains non-convex akin to $(\mathrm{P}_2)$. Problem $(\mathrm{P}_3)$ is homogenized as
\vspace{0pt}
\begin{align}
  \mathrm{P_{3}^\prime}: \quad & \max_{\vs\in\R^6, t\in\R} \bvs^\top \bar{\mathbf{Q}}_0 \bvs \\
  \mathrm{s.t.} \quad & \bvs^\top \bar{\mathbf{Q}}_i \bvs \leq 1, \ i = 1,2,3,4,5, \quad t^2 = 1, \nonumber \\
  \quad & \bvs^\top \bar{\mathbf{Q}}_i \bvs = 0, \ i = 6,7,8, \quad \bvs = \begin{bmatrix} \vs^\top &  t \end{bmatrix}^\top, \nonumber \\
  \mathrm{with} \quad & \bar{\mathbf{Q}}_0 = \begin{bmatrix}
    \mathbf{Q}_0 & \mathbf{c} \\
    \mathbf{c}^\top & 0
  \end{bmatrix}, \ \bar{\mathbf{Q}}_i = \frac{1}{V_i} \cdot \begin{bmatrix}
    \mathbf{Q}_i & \\
    & 0
  \end{bmatrix}\succcurlyeq\zero, \ i = 1,2, \nonumber \\
      \quad & \bar{\mathbf{Q}}_i = \frac{1}{A_{i-2}} \cdot \begin{bmatrix}
        \mathbf{Q}_{i-2} & -\mathbf{b}_{i-2} \\
        -\mathbf{b}_{i-2}^\top & \|\mathbf{b}_{i-2}\|^2
      \end{bmatrix}\succcurlyeq\zero, \quad i = 3,4, \nonumber \\
        \quad & \bar{\mathbf{Q}}_5 = \frac{1}{V^2}\cdot\begin{bmatrix}
          \mathbf{Q}_{3} & -\Smat^\top\Rmat_1^\top\Rmat_2\mathbf{b}_3 \\
          -\mathbf{b}_3^\top\Rmat_2^\top\Rmat_1\Smat & \|\mathbf{b}_3\|^2
        \end{bmatrix}, \nonumber \\
    \quad & \bar{\mathbf{Q}}_i = \begin{bmatrix}
      \mathbf{Q}_{i-2} & \\
    & 0
    \end{bmatrix}\succcurlyeq\zero, \qquad i = 6,7,8, \nonumber
\end{align} 
where $\bar{\mathbf{Q}}_0\in\mathbb{S}^7$, and the positive semidefiniteness of $\bar{\mathbf{Q}}_i$, for $i = 3,4$, is proven via the Theorem 1.20 in \cite{zhang2006schur}: (i) $\mathbf{Q}_{i-2}\succcurlyeq\zero$; (ii) $\mathbf{b}_{i-2}\in\mathcal{R}(\mathbf{Q}_{i-2})$; (iii) $\|\mathbf{b}_{i-2}\|^2-\mathbf{b}_{i-2}^\top\mathbf{Q}_{i-2}^{\dagger}\mathbf{b}_{i-2}=0$, thus, $\bar{\mathbf{Q}}_i / \mathbf{Q}_{i-2}\succcurlyeq\zero$. The same applied to $\bar{\mathbf{Q}}_5$, assuming $\Rmat_1\approx\Rmat_2$. The semidefinite relaxation (SDR) is applied and solved using convex solvers, e.g., CVX \cite{grant2014cvx}. Subsequently, a rank-one approximation is extracted using Gaussian randomization.
\begin{rem}
  The objective function resembles a local \textit{cost-to-go} function for an elementary transition in MPC, aiming to minimize the total cost over a finite horizon. However, applying this analogy faces several challenges: (i) There is no well-defined reference to use control $\boldsymbol{\xi}^\star$ to maintaining an equilibrium state $\T^\star$. (ii) The objective function exhibits a highly nonlinear dependence on the state through $\mathbf{c}$ and $\bar{\mathbf{P}}$, rather than being merely the distance to the reference with some chosen metrics. (iii) This local approach may not align well with the global optimal trajectory over the long term.
\end{rem}
%
%
%

\section{Numerical Results}
This section analyzes the performance of target tracking and the trajectory generation via control signal design. Key simulation parameters are: $N_T = N_R = 2 \times 2 = 4$, $f_c = 2.4$ GHz, $L=32$, $K=10$, $\sigmarcs=0.5$ m, $\mathbf{s}=[0.2,0.3,0.1]^\top$ m, $\Dt = 0.25$ s, $N_e = 200$, $V_{\ell} = 6$ m/s, $V_{a} = 0.15$ rad/s, $A_{\ell} = 2$m/s, $A_{a} = 0.05$ rad/s, $V = 0.5$ m/s. The GU is initialized as $\Twp^0=(\I_3,[200, 150, 0]^\top\text{m})$, i.e., its axes align with the global frame. It moves with consitant velocity 4m/s, along the x-axis from right to left. The UAV begins at $\Tws^0=(\diag(-1,1,-1),[200, 0, 150]^\top\text{m})$, i.e., with its axises rotated \ang{180} about the y-axis, with the inital control $^{\mathcal{B}}\boldsymbol{\xi}_s^0 = \left[0, 1, \zero_4^\top\right]^\top$. The EKF is initialized via a coarse estimate of the GU position $\hat{\rvec}_{w,p}^0=\left[200, 170, 0\right]^\top\text{m}$. 

\begin{figure}
  \centering
  \begin{tabular}{ccc}
    \epsfig{file=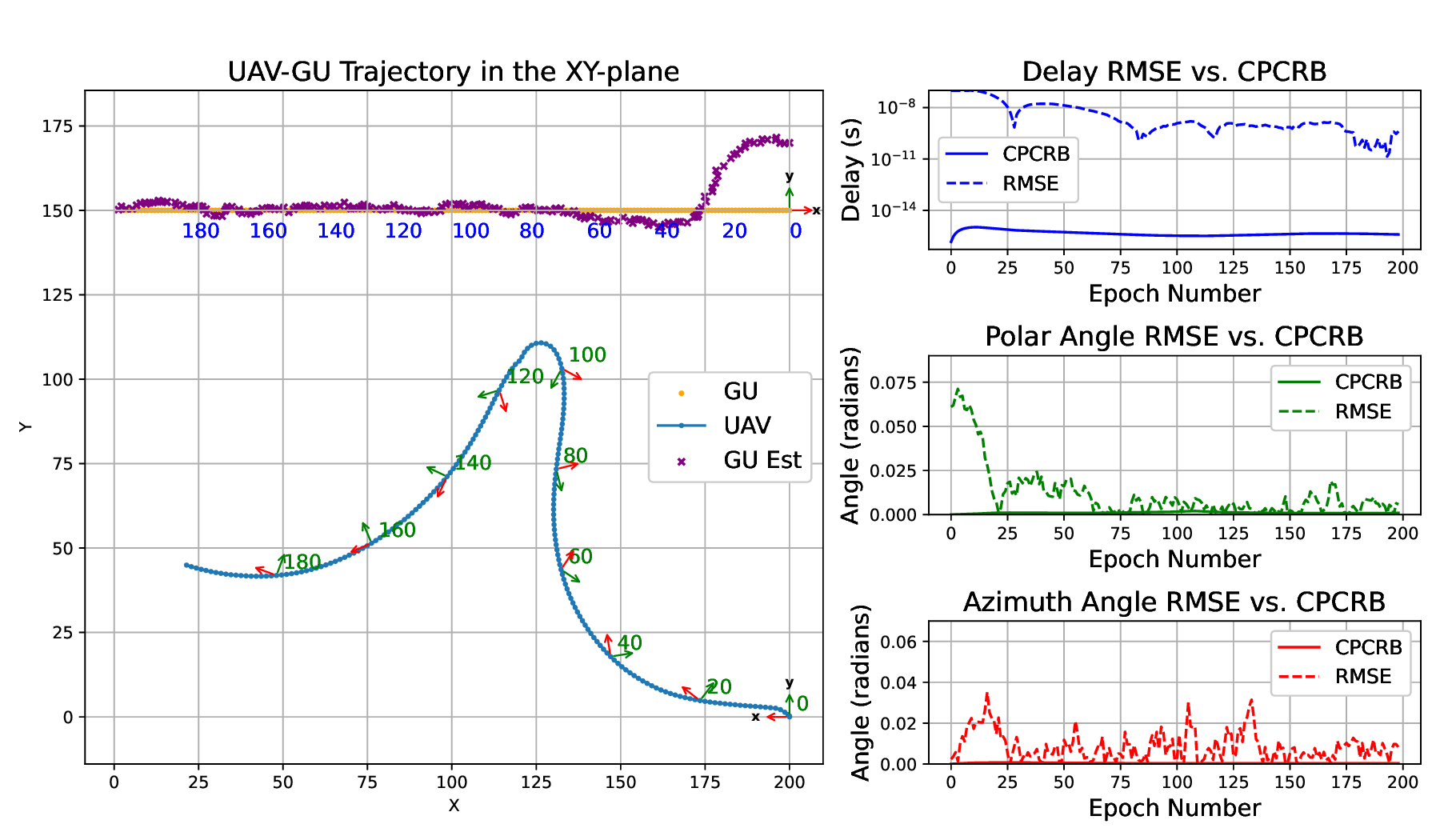, width=\linewidth, clip=} \\
    \epsfig{file=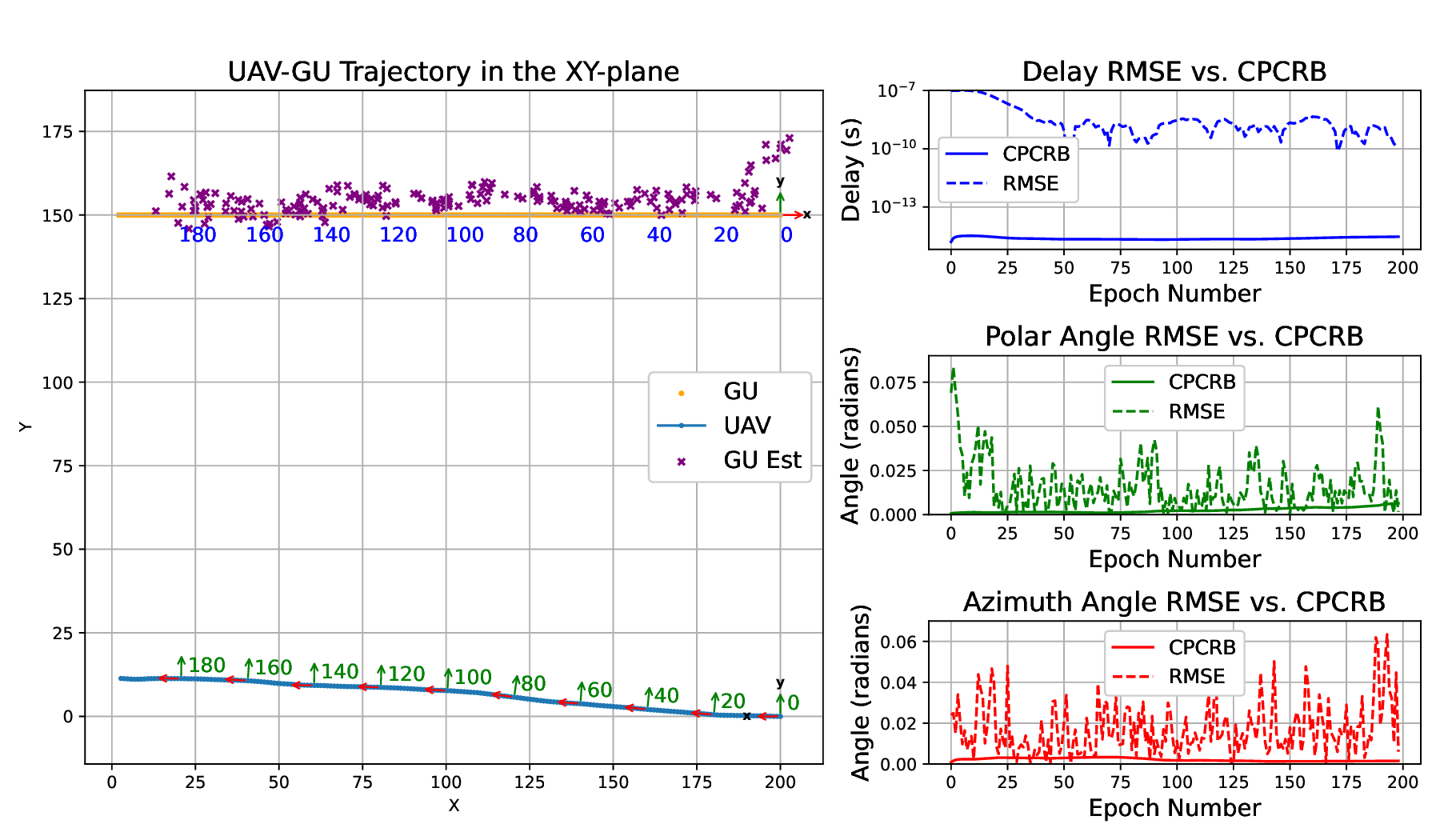, width=\linewidth, clip=} \\
    \epsfig{file=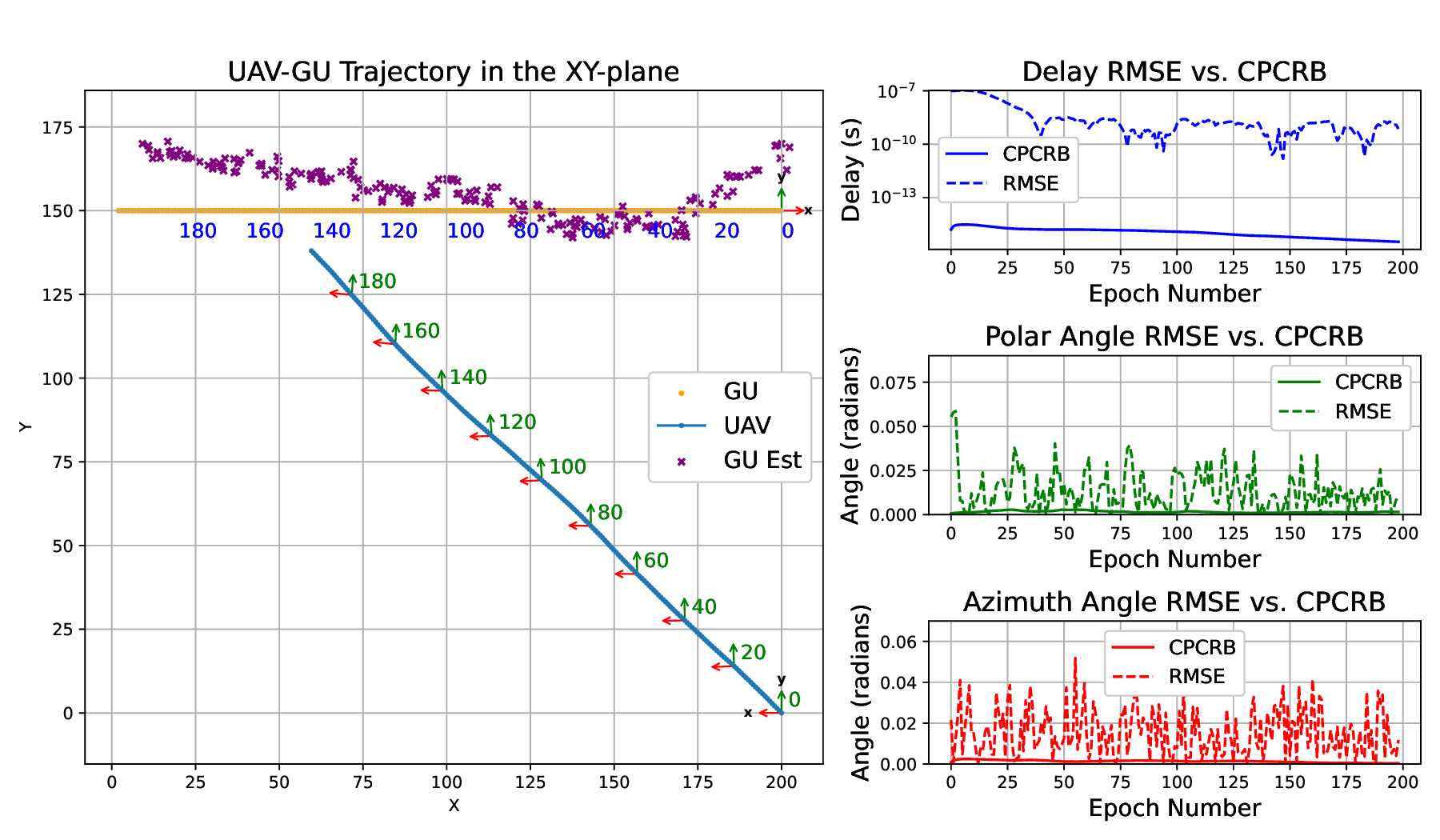, width=\linewidth, clip=} 
  \end{tabular}
  \caption{UAV trajectories and the RMSE vs. CPCRB metrics for different settings: (top) optimised, (middle) parallel, and (bottom) diagonal settings.}
  \label{fig.1}
\end{figure}

We evaluate radar sensing performance in terms of root mean square error (RMSE) for delay, polar and azimuthal angles—key metrics in radar applications—by comparing three trajectory designs in Fig. \ref{fig.1}: (i) an optimized trajectory by solving $(\mathrm{P_{3}^\prime})$ (top), (ii) a parallel trajectory (middle) aligned with the GU's path, and (iii) a diagonal trajectory (bottom), approaching the GU at a \ang{45}-angle with the same velocity in (ii). First, we examine the GU position estimates for different trajectories, shown in the left panel of Fig. \ref{fig.1}: The optimized trajectory achieves better convergence in GU position estimates. In the parallel setting, the UAV deviates slightly from the GU's path due to control noise $\boldsymbol{\xi}_{\mathbf{w}}$, and the GU position estimates show acceptable accuracy but low precision. In the diagonal trajectory, as the UAV approaches the GU, the position estimates exhibit significant systematic deviations from the true values, i.e., with high precision but low accuracy. This might be attributed to the increasing relative velocity between the UAV and the GU, limiting the EKF's ability to adapt quickly to rapid changes. A similar phenomenon is observed in \cite{liu2020radar}, where RMSE spikes occur, despite a decreasing PCRB, as the vehicle passes in front of the radar. Secondly, we evaluate the RMSE vs. CPCRB metrics w.r.t. trajectories, shown in the right panel of Fig. \ref{fig.1}. The optimized trajectory outperforms the heuristic designs, with RMSE decreasing steadily and stabilizing close to the CRLB (RMSE of both angles stays below 0.02 rad), despite: (i) a stabilizing stage in delay estimate between 25-75 epochs; (ii) requiring more epochs (>25) to converge to the true parameter values compared to the heuristic designs. 
\section{Conclusion}
This paper develops a novel UAV-enabled MIMO-OFDM ISAC system that incorporates UPA configurations and rigid-body dynamics via $SE(3)$, enabling both target tracking and trajectory design from a control perspective. It highlights that the performance of target tracking and parameter estimation depends not only on the geometric relationship but also on the dynamic changes within the system, underscoring the critical role of UAV trajectory design in enhancing radar sensing performance.


\section{Appendix A}
\subsection{Derivation of Relative Velocity for Doppler Frequency Shift} \label{sub.1.4}
To calculate the Doppler frequency shift by virtue of the UAV and GU mobility, we first examine their velocities in the global frame $\{w\}$. For the UAV, $\mathbf{s}\in\R^3$ represents the midpoint coordinates of the UPA, relative to the UAV's rotation center, with its representation in \textit{homogeneous coordinates} \cite{gallo2022so}, i.e., $\bar{\mathbf{s}}=\big[\mathbf{s}^\top, \ 1\big]^\top\in\mathbb{R}^4$. These coordinates are determined by the UPA installation on the UAV platform. The global coordinates are given as $^{\mathcal{E}}\bar{\mathbf{s}}_{n}=\mathbf{T}_{w,s}^{n}\bar{\mathbf{s}}$. By taking the time direvative, we obtain its \textit{global velocity}, i.e., 
\begin{equation}
 	^{\mathcal{E}}\bar{\mathbf{v}}_{s}^n\triangleq{^{\mathcal{E}}\dot{\bar{\mathbf{s}}}_{n}}=\dot{\mathbf{T}}_{w,s}^{n}\bar{\mathbf{s}}=\mathbf{T}_{w,s}^{n}\big({\vsB^{n+1}}\big)^{\wedge}\bar{\mathbf{s}}\triangleq\mathbf{T}_{w,s}^{n}{\velosB^n},
\end{equation}
where the property $\big({\vsB^{n+1}}\big)^{\wedge}=\big(\mathbf{T}_{w,s}^{n}\big)^{-1}\dot{\mathbf{T}}_{w,s}^{n}$ is utilised \cite{gallo2022so}, and the \textit{local velocity} reads as ${\velosB^n}\triangleq\big({\vsB^{n+1}}\big)^{\wedge}\bar{\mathbf{s}}=[^{\mathcal{B}}\mathbf{v}_s^{n,\top}, 0]^\top$. For the GU, as a single-point target, we consider only the origin of $\{p_n\}$, with its homogeneous coordinates $\bar{\mathbf{p}}=\big[\mathbf{0}^\top, \ 1\big]^\top\in\mathbb{R}^4$. We have its global velocity, given as
\begin{equation}
  ^{\mathcal{E}}\bar{\mathbf{v}}_{p}^n=\mathbf{T}_{w,p}^{n}\big( {\vpB^{n+1}}\big)^{\wedge}\bar{\mathbf{p}}=\mathbf{T}_{w,p}^{n}{^{\mathcal{B}}\bar{\mathbf{v}}_{p}^n},
\end{equation}
where the local velocity reads as $^{\mathcal{B}}\bar{\mathbf{v}}_{p}^n\triangleq\big( {\vpB^{n+1}}\big)^{\wedge}\bar{\mathbf{p}}=[^{\mathcal{B}}\mathbf{v}_p^{n,\top}, 0]^\top$. Therefore, the \textit{relative velocity vector} in the global frame $\{w\}$ is computed as 
\begin{equation}
  ^{\mathcal{E}}\bar{\mathbf{v}}_{s,p}^n={^{\mathcal{E}}\bar{\mathbf{v}}_{p}^n} - {^{\mathcal{E}}\bar{\mathbf{v}}_{s}^n}.
\end{equation}
Additionally, the \textit{position vector}, in its homogeneous coordinates \cite{gallo2022so}, is expressed as $\bar{\mathbf{r}}_{s, p}^n = \left[\mathbf{r}_{s, p}^{n,\top}, 0\right]^\top\in\mathbb{R}^4$ locally in $\{s_n\}$, with its global representation acquired by pre-multiplying $\mathbf{T}_{w, s}^n$, i.e., $^{\mathcal{E}}\bar{\mathbf{r}}_{s, p}^n=\mathbf{T}_{w, s}^n\bar{\mathbf{r}}_{s, p}^n$. Therefore, the \textit{relative velocity} of $\{p_n\}$ w.r.t. $\{s_n\}$ is given as 
\begin{equation} \label{eq.1.77}
 \begin{aligned}
   v_{\text{radial}}^n &= \left\langle^{\mathcal{E}}\bar{\mathbf{v}}_{s,p}^n,^{\mathcal{E}}\bar{\mathbf{r}}_{s, p}^n\right\rangle / \rho_{s,p}^n \\
  &\stackrel{(a)}{=}\left\langle\Tsw^n{^{\mathcal{E}}\bar{\mathbf{v}}_{s,p}^n},\Tsw^n{^{\mathcal{E}}\bar{\mathbf{r}}_{s, p}^n}\right\rangle / \rho_{s,p}^n \\
  &=\left\langle\Tsp^n{{\velopB^n}}-{{\velosB^n}},{\bar{\mathbf{r}}_{s, p}^n}\right\rangle / \rho_{s,p}^n \\
  &=\left\langle\mathbf{R}_{s,p}^n{^{\mathcal{B}} \mathbf{v}_p^n}-{^{\mathcal{B}} \mathbf{v}_s^n}, \mathbf{r}_{s,p}^n\right\rangle / \rho_{s,p}^n,
 \end{aligned} 
\end{equation}
where the property -- that a rigid-body motion preserves the inner product of vectors \cite{ma2004invitation} -- is applied in $(a)$.

\subsection{Computation of Jacobian block $\J_{\boldsymbol{\zeta}}^{\mathbf{h}}$} \label{sub.1.1}
The dependence of the transmit and receive array response vectors on polar and azimuthal angles is omitted for simplicity, i.e., $\mathbf{a}_{T}\triangleq\mathbf{a}_{T}(\theta, \phi)$, $\mathbf{a}_{R}\triangleq\mathbf{a}_{R}(\theta, \phi)$. 
\paragraph{$\mathbf{j}^{\mathbf{h}}_{\tau}$}
Considering (\ref{eq.1.36}) and (\ref{eq.1.43}), there exist an invertible function between $\tau\in\R_+$ and $\left|b\right| \in\R_+$, allowing $\left|b\right|$ to be fully expressed in terms of $\tau$. Therefore,
\begin{equation} \label{eq.1.55}
	\begin{split}
		&\mathbf{j}^{\mathbf{h}}_{\tau}=b^\prime\cdot\boldsymbol{\omegatilde}\otimes\mathbf{a}_{T}^* \otimes \mathbf{a}_{R} + b\cdot\mathbf{j}^{\boldsymbol{\omegatilde}}_{\tau}\otimes\mathbf{a}_{T}^* \otimes \mathbf{a}_{R} \\
		&=-\frac{2b}{\tau}\left(\boldsymbol{\omegatilde}\otimes\mathbf{a}_{T}^* \otimes \mathbf{a}_{R}\right)-j2\pi b f_0\cdot \left(\boldsymbol{\omegatilde}\otimes\mathbf{a}_{T}^* \otimes \mathbf{a}_{R}\right)\odot\boldsymbol{\ell} \\
    &\triangleq-j2\pi b f_0\cdot \left(\boldsymbol{\omegatilde}\otimes\mathbf{a}_{T}^* \otimes \mathbf{a}_{R}\right)\odot\boldsymbol{\tilde{\ell}},
	\end{split}
\end{equation}
where $\mathbb{R}^{MN_TN_R}\ni\boldsymbol{\ell}\triangleq\boldsymbol{\ell}_0\otimes\ones_{N_TN_R}$ is a constant and deterministic vector with $\boldsymbol{\ell}_0\triangleq\left[\ell_1,\ldots,\ell_M\right]^\top$ which only depends on the RE and UPA configuration, operator $\odot$ is the Hadamard product and the property $(\mathbf{A} \odot \mathbf{C}) \otimes(\mathbf{B} \odot \mathbf{D})=(\mathbf{A} \otimes \mathbf{B}) \odot(\mathbf{C} \otimes \mathbf{D})$ is applied, assuming $\mathbf{A}$ has the same dimension as $\mathbf{C}$ and $\mathbf{B}$ has the same dimension as $\mathbf{D}$, thus, $(\mathbf{a} \odot \mathbf{c}) \otimes \mathbf{b}=(\mathbf{a} \otimes \mathbf{b}) \odot(\mathbf{c} \otimes \mathbf{1})$.
\paragraph{$\mathbf{j}^{\mathbf{h}}_{\mu}$}
Following the same procedure in (\ref{eq.1.55}), we give the resultant Jacobian directly, i.e.,
\begin{equation} \label{eq.1.56}
	\mathbf{j}^{\mathbf{h}}_{\mu}= j2\pi b T_s\cdot \left(\boldsymbol{\omegatilde}\otimes\mathbf{a}_{T}^* \otimes \mathbf{a}_{R}\right)\odot\mathbf{k},
\end{equation}
where $\mathbb{R}^{MN_TN_R}\ni\mathbf{k}\triangleq\mathbf{k}_0\otimes\ones_{N_TN_R}$ is a constant and deterministic vector with $\mathbf{k}_0\triangleq\left[k_1,\ldots,k_M\right]^\top$ which only depends on the RE and UPA configuration,
\paragraph{$\mathbf{j}^{\mathbf{h}}_{\phi}$}
Hereby, we give its derivation as follows
\begin{align}
	\mathbf{j}_\phi^{\mathbf{h}} &= b \cdot \boldsymbol{\omegatilde} \otimes \left[ \mathbf{j}_\phi^{\mathbf{a}_{T,y}^*} \otimes \mathbf{a}_{T,x}^* \otimes \mathbf{a}_{R,y} \otimes \mathbf{a}_{R,x} \right. \label{eq.1.50} \\
	&\quad + \mathbf{a}_{T,y}^* \otimes \mathbf{j}_\phi^{\mathbf{a}_{T,x}^*} \otimes \mathbf{a}_{R,y} \otimes \mathbf{a}_{R,x} \label{eq.1.47} \\
	&\quad + \mathbf{a}_{T,y}^* \otimes \mathbf{a}_{T,x}^* \otimes \mathbf{j}_\phi^{\mathbf{a}_{R,y}} \otimes \mathbf{a}_{R,x} \label{eq.1.48} \\
	&\quad \left. + \mathbf{a}_{T,y}^* \otimes \mathbf{a}_{T,x}^* \otimes \mathbf{a}_{R,y} \otimes \mathbf{j}_\phi^{\mathbf{a}_{R,x}} \right]. \label{eq.1.49}
\end{align}
The computation for the first term in the brackets is given as
\begin{equation*}
	\begin{split}
		&\text{(\ref{eq.1.50}): }\quad\mathbf{j}_\phi^{\mathbf{a}_{T,y}^*} \otimes \mathbf{a}_{T,x}^* \otimes \mathbf{a}_{R,y} \otimes \mathbf{a}_{R,x} \\
    &= j\pi\sin\theta\cos\phi\cdot\left(\aTy^*\odot\nTy\right)\otimes\aTx^*\otimes \mathbf{a}_{R,y} \otimes \mathbf{a}_{R,x} \\
		&= j\pi\sin\theta\cos\phi\cdot\left(\aT^*\otimes\aR\right)\odot\left(\nTy\otimes\ones\otimes\ones\otimes\ones\right),
	\end{split}
\end{equation*}
The remaining three terms can be derived straightforwardly and are directly given as follows
\begin{alignat*}{3}
	&\text{(\ref{eq.1.47}): } \quad -j\pi\sin\theta\sin\phi\cdot\left(\aT^*\otimes\aR\right)&&\odot\left(\ones\otimes\nTx\otimes\ones\otimes\ones\right), \\
	&\text{(\ref{eq.1.48}): } \quad -j\pi\sin\theta\cos\phi\cdot\left(\aT^*\otimes\aR\right)&&\odot\left(\ones\otimes\ones\otimes\nRy\otimes\ones\right), \\
	&\text{(\ref{eq.1.49}): } \quad \quad j\pi\sin\theta\sin\phi\cdot\left(\aT^*\otimes\aR\right)&&\odot\left(\ones\otimes\ones\otimes\ones\otimes\nRx\right),
\end{alignat*}
\begin{equation*}
	\resizebox{\hsize}{!}{$
	\begin{aligned}
		\nTx & \triangleq \frac{1}{\sqrt{N_{T,x}}}\left[0,\ldots,N_{T,x}-1\right]^\top, 
		\nTy \triangleq \frac{1}{\sqrt{N_{T,y}}}\left[0,\ldots,N_{T,y}-1\right]^\top, \\
		\nRx & \triangleq \frac{1}{\sqrt{N_{R,x}}}\left[0,\ldots,N_{R,x}-1\right]^\top, 
		\nRy  \triangleq \frac{1}{\sqrt{N_{R,y}}}\left[0,\ldots,N_{R,y}-1\right]^\top.
	\end{aligned}
	$}
\end{equation*}
Therefore, we have
\begin{equation} \label{eq.1.51}
	\resizebox{\hsize}{!}{$
		\begin{aligned}
			&\mathbf{j}_\phi^{\mathbf{h}} = j\pi b\sin\theta \cdot \boldsymbol{\omegatilde} \otimes  \Big[ \left(\aT^* \otimes \aR\right) \odot \\
			& \Big( 
			\cos\phi \left(\nTy \otimes \ones \otimes \ones \otimes \ones\right) - \sin\phi \left(\ones \otimes \nTx \otimes \ones \otimes \ones\right) \\
			& - \cos\phi \left(\ones \otimes \ones \otimes \nRy \otimes \ones\right) + \sin\phi \left(\ones \otimes \ones \otimes \ones \otimes \nRx\right) \Big) \Big] \\
			&= j\pi b\sin\theta \cdot \boldsymbol{\omegatilde} \otimes\left(\aT^* \otimes \aR\right) \odot \\
			&\Big[ \cos\phi \left(\ones \otimes\nTy \otimes \ones \otimes \ones \otimes \ones\right) - \sin\phi \left(\ones \otimes\ones \otimes \nTx \otimes \ones \otimes \ones\right) \\
			& - \cos\phi \left(\ones \otimes\ones \otimes \ones \otimes \nRy \otimes \ones\right) + \sin\phi \left(\ones \otimes\ones \otimes \ones \otimes \ones \otimes \nRx\right) \Big) \Big] \\
			&=  j\pi b\sin\theta \cdot \left(\boldsymbol{\omegatilde} \otimes\aT^* \otimes \aR\right) \odot \left(\mathbf{N}_1\cdot\mathbf{f}_\phi\right),
		\end{aligned}
		$}
\end{equation}
where the property $\mathbf{a} \otimes(\mathbf{b} \odot \mathbf{d})=(\mathbf{a} \otimes \mathbf{b}) \odot(\mathbf{1} \otimes \mathbf{d})$ is applied, with $\mathbf{N}_1\in\mathbb{R}^{MN_TN_R\times 4}$ as a constant and deterministic matrix, which only depends on the RE and UPA configurations, i.e.,
\begin{equation}
	\resizebox{0.87\hsize}{!}{$
		\mathbf{N}_1\triangleq\begin{bmatrix}
			\left(\ones_{M} \otimes\nTy \otimes \ones_{N_{T,x}} \otimes \ones_{N_{R,y}} \otimes \ones_{N_{R,x}}\right)^\top \\
			\left(\ones_{M} \otimes\ones_{N_{T,y}} \otimes \nTx \otimes \ones_{N_{R,y}} \otimes \ones_{N_{R,x}}\right)^\top \\
			\left(\ones_{M} \otimes\ones_{N_{T,y}} \otimes \ones_{N_{T,x}} \otimes \nRy \otimes \ones_{N_{R,x}}\right)^\top \\
			\left(\ones_{M} \otimes\ones_{N_{T,y}} \otimes \ones_{N_{T,x}} \otimes \ones_{N_{R,y}} \otimes \nRx\right)^\top \\
		\end{bmatrix}^\top,
		$}
\end{equation}
where the dimension of each vector $\ones$ is specified across (\ref{eq.1.50})-(\ref{eq.1.51}), and the $\mathbf{f}_\phi$ as a function of polar angle is given as $\mathbf{f}_\phi = \begin{bmatrix}
	\cos\phi \ -\sin\phi \ -\cos\phi \ \sin\phi
\end{bmatrix}^\top$.
\paragraph{$\mathbf{j}^{\mathbf{h}}_{\theta}$}
Following the similar process from (\ref{eq.1.50})-(\ref{eq.1.51}), we obtain
\begin{equation} \label{eq.1.54}
	\resizebox{\hsize}{!}{$
		\begin{aligned}
			&\mathbf{j}_\theta^{\mathbf{h}} =  j\pi b\cos\theta \cdot \boldsymbol{\omegatilde} \otimes\left(\aT^* \otimes \aR\right) \odot \\
			&\Big[ \sin\phi \left(\ones \otimes\nTy \otimes \ones \otimes \ones \otimes \ones\right) + \cos\phi \left(\ones \otimes\ones \otimes \nTx \otimes \ones \otimes \ones\right) \\
			& - \sin\phi \left(\ones \otimes\ones \otimes \ones \otimes \nRy \otimes \ones\right) - \cos\phi \left(\ones \otimes\ones \otimes \ones \otimes \ones \otimes \nRx\right) \Big) \Big] \\
			&= j\pi b\cos\theta \cdot \left(\boldsymbol{\omegatilde} \otimes\aT^* \otimes \aR\right) \odot \left(\mathbf{N}_2\cdot\mathbf{f}_\phi\right),
		\end{aligned}
		$}
\end{equation}
with $\mathbf{N}_2\in\mathbb{R}^{MN_TN_R\times 4}$ as another constant and deterministic matrix, which can be derived from $\mathbf{N}_1$, i.e.,
\begin{equation}
	\mathbf{N}_2 \triangleq \mathbf{N}_1\cdot\begin{bmatrix}
    0 & -1 & 0 & 0 \\
    0 & 0 & -1 & 0 \\
    0 & 0 & 0 & -1 \\
    1 & 0 & 0 & 0
	\end{bmatrix}.
\end{equation}
\subsection{Computation of Jacobian block $\J_\T^{\boldsymbol{\zeta}}$} \label{sub.1.0}
\paragraph{$\mathbf{J}^{\tau}_{\mathbf{T}}$}
Considering (\ref{eq.1.36}) and (\ref{eq.1.77}), the dependence of delay $\tau$ on $\mathbf{T}$ is explicitly denoted as
\begin{equation}
	\tau = \frac{2}{c}\|\mathbf{r}\|=\frac{2}{c}\|g_\T(\zero)\|, \label{eq.1.46}
\end{equation}
where $g_{\mathbf{T}}(\cdot)$ denotes the \textit{homogeneous representations} of motion actions \cite{gallo2022so}. Therefore, we have \begin{equation}
	\begin{split}
		\J_\T^\tau&=\frac{\partial \tau}{\partial \T}=\frac{2}{c}\frac{\partial \|\mathbf{r}\|}{\partial \mathbf{r}}\frac{\partial \mathbf{r}}{\partial \T}=\frac{2}{c}\frac{\rvec^\top}{\rnorm}\J_\T^{g_\T(\zero)}\\
		&=\frac{2}{c}\frac{\rvec^\top}{\rnorm}\begin{bmatrix}
			\mathbf{R} & -\mathbf{R} \left[{\mathbf{0}}\right]_\times
		\end{bmatrix} =\frac{2}{c}\frac{\rvec^\top}{\rnorm}\begin{bmatrix}
			\mathbf{R} & \mathbf{0}_{3\times 3}
		\end{bmatrix},
	\end{split}
\end{equation}
where the basic Jacobian blocks in $SE(3)$ are utilized, cf. Table 5.8 of \cite{gallo2022so}, and the operator $\left[\cdot\right]_\times$ denotes the \textit{skew-symmetric matrix} representation of a vector \cite{mueller2019modern}.
\paragraph{$\mathbf{J}^{\mu}_{\mathbf{T}}$}
Considering (\ref{eq.1.33}), the dependence of delay $\mu$ on $\mathbf{T}$ is explicitly denoted as
\begin{equation}
	\begin{aligned}
		\mu&=\frac{2f_c}{c\rnorm}\langle \T{\velopB} - {\velosB}, \rbar\rangle = \frac{2f_c}{c\rnorm}\langle \mathbf{R}{^{\mathcal{B}}{\mathbf{v}}_{p}} - {^{\mathcal{B}}{\mathbf{v}}_{s}}, \rvec\rangle\\
		&=\frac{2f_c}{c\rnorm}(\langle g_{\T^*}({\velopB})-{\velosB}, g_{\T}(\bar{\zero})-\bar{\zero} \rangle) \\
		&=\frac{2f_c}{c\rnorm}(\langle g_{\T^*}({\velopB})-{\velosB}, g_{\T}(\bar{\zero}) \rangle) \\
		&=\frac{2f_c}{c\rnorm}(\langle g_{\T^*}({\velopB}), g_{\T}(\bar{\zero}) \rangle -\langle{^{\mathcal{B}}{\mathbf{v}}_{s}}, g_{\T}({\zero}) \rangle),
	\end{aligned}
\end{equation}
where $g_{\mathbf{T}}(\cdot)$, as the homogeneous representations of motion actions, shares the same expression on points and vectors, i.e., $g_{\mathbf{T}^{*}}(\overline{\mathbf{v}})=g_{\mathbf{T}}(\overline{\mathbf{v}})$, cf. Section 5.3 of \cite{gallo2022so}, with $\overline{\mathbf{v}}$ as the \textit{homogeneous coordinates} of a vector $\mathbf{v} \in \mathbb{R}^3$, i.e., $\overline{\mathbf{v}}=\left[\mathbf{v}^{\top}, 0\right]^{\top} \in \mathbb{R}^4$. The intermediate Jacobian blocks for computation of $\mathbf{J}^{\mu}_{\mathbf{T}}$ are given as follows
\begin{equation} \label{eq.1.74}
	\frac{\partial \|\mathbf{r}\|^{-1}}{\partial \mathbf{T}} = \frac{\partial \|\mathbf{r}\|^{-1}}{\partial \mathbf{r}} \frac{\partial \mathbf{r}}{\partial \mathbf{T}}=-\rnorm^{-3}\cdot\mathbf{r}^\top\begin{bmatrix}
		\mathbf{R} & \mathbf{0}_{3\times 3}
	\end{bmatrix},
\end{equation}
\begin{equation} \label{eq.1.75}
	\frac{\partial}{\partial \T}\langle{^{\mathcal{B}}{\mathbf{v}}_{s}}, g_{\T}({\zero}) \rangle={^{\mathcal{B}}{\mathbf{v}}_{s}^\top}\J_\T^{g_\T(\zero)}={^{\mathcal{B}}{\mathbf{v}}_{s}^\top}\begin{bmatrix}
		\mathbf{R} & \mathbf{0}_{3\times 3}
	\end{bmatrix},
\end{equation}
\begin{equation} \label{eq.1.57}
	\frac{\partial}{\partial \T}\langle g_{\T^*}({\velopB}), g_{\T}(\bar{\zero}) \rangle=\begin{bmatrix}
		{^{\mathcal{B}}{\mathbf{v}}_{p}^\top} & -\mathbf{r}^\top\mathbf{R}\left[^{\mathcal{B}}{\mathbf{v}}_{p}\right]_\times
	\end{bmatrix},
\end{equation}
Therefore, we have
\begin{equation}
	\begin{split}
		\mathbf{J}^{\mu}_{\mathbf{T}}=&\frac{2f_c}{c\rnorm}\left(
		\begin{bmatrix}
			{^{\mathcal{B}}{\mathbf{v}}_{p}^\top}-{^{\mathcal{B}}{\mathbf{v}}_{s}^\top}\mathbf{R}
			& -\rvec^\top\mathbf{R}\left[{^{\mathcal{B}}{\mathbf{v}}_{p}}\right]_{\times}	\end{bmatrix} \right.\\
		&\left. -\rnorm^{-2}\cdot\langle \mathbf{R}{^{\mathcal{B}}{\mathbf{v}}_{p}} - {^{\mathcal{B}}{\mathbf{v}}_{s}}, \rvec\rangle\cdot\begin{bmatrix}
			\rvec^\top\mathbf{R} & \zero_{1\times 3}
		\end{bmatrix}
		\right).
	\end{split}
\end{equation}
\begin{proof}
  The proofs of \eqref{eq.1.74} and \eqref{eq.1.75} are straightforward, while the proof of (\ref{eq.1.57}) is given in the sequel. Let $f: SE(3)\to\R, \T\mapsto f(\T)=\langle g_{\T^*}(\bar{\mathbf{v}}), g_{\T}(\bar{\zero}) \rangle$, where $\bar{\zero}$ is the homogeneous representation of the origin, i.e, $\efour=\bar{\mathbf{0}}=[\mathbf{0}^\top, 1]^\top\in\R^4$. By taking the \textit{directional derivative} of $f(\T)$ along the direction $\boldsymbol{\tau}=[\rhovec^\top, \thetavec^\top]^\top$ in the local tangent space, we have
\begin{align}
	&\D_{\tauvec}f(\T) \nonumber \\
	=&\lim_{t\to 0}\frac{\langle \left( \T\oplus t\tauvec \right)\vbar,\left( \T\oplus t\tauvec \right)\efour\rangle-\langle \T\vbar,\T\efour\rangle}{t} \nonumber \\
	=&\lim_{t\to 0}\frac{\langle  \T\cdot\Exp(t\tauvec) \cdot \vbar, \T\cdot\Exp(t\tauvec) \cdot \efour\rangle-\langle \T\vbar,\T\efour\rangle}{t} \nonumber \\
	\approx&\lim_{t\to 0}\frac{\langle \mathbf{R}(\I+t\left[\thetavec\right]_\times)\mathbf{v}, t\mathbf{R}\rhovec+\rvec \rangle - \langle \mathbf{Rv},\rvec \rangle}{t} \nonumber \\
	=&\lim_{t\to 0}\frac{\langle \mathbf{Rv}, t\mathbf{R}\rhovec \rangle+\langle t\mathbf{R}\left[\thetavec\right]_\times\mathbf{v}, t\mathbf{R}\rhovec \rangle+\langle t\mathbf{R}\left[\thetavec\right]_\times\mathbf{v}, \rvec \rangle}{t} \nonumber \\
	=& \mathbf{v}^\top\rhovec+\mathbf{r}^\top\mathbf{R}\left[\thetavec\right]_\times\mathbf{v}
	=\begin{bmatrix}
		\mathbf{v}^\top & -\rvec^\top\mathbf{R}\left[\mathbf{v}\right]_\times
	\end{bmatrix}\cdot \tauvec,
\end{align}
where the property $[\mathbf{a}]_{\times} \mathbf{b}=-[\mathbf{b}]_{\times}\mathbf{a}$ is applied and the appromixation used here is given as \cite{sola2021micro}
\begin{equation}
	\Exp(t\tauvec)\approx\begin{bmatrix}
		\I+t\left[\thetavec\right]_\times & t\rhovec \\
		\zero & 1
	\end{bmatrix}.
\end{equation}
Therefore, for the given $f(\T)$, the \textit{differential map} $\D f(\T)$ has a matrix form and the associated composition corresponds to the matrix-vector multiplication, i.e., $\D f(\T)=\frac{\partial f(\T)}{\partial \T}=\begin{bmatrix}
	\mathbf{v}^\top & -\rvec^\top\mathbf{R}\left[\mathbf{v}\right]_\times
\end{bmatrix}$.
\end{proof}

\paragraph{$\mathbf{J}^{\theta}_{\mathbf{T}}$}
Considering (\ref{eq.1.37}), the dependence of delay $\theta$ on $\mathbf{T}$ is explicitly denoted as
\begin{equation}
	\begin{split}
		\theta&=\operatorname{accos}\left(\frac{z}{\rnorm}\right)=\operatorname{accos}\left(\frac{\langle \ethree,\rvec \rangle}{\rnorm}\right)\\
		&=\operatorname{accos}\left(\frac{\langle \ethree,g_\T(\zero) \rangle}{\rnorm}\right),
	\end{split}
\end{equation}
where $\ethree=[0,0,1]^\top\in\R^3$ is used to extract the element $z$ from $\rvec$. The imtermediate Jacobian block for computation of $\mathbf{J}^{\theta}_{\mathbf{T}}$ are given as follows
\begin{equation}
	\begin{split}
		&\frac{\partial}{\partial\T}\frac{\langle \ethree,g_\T(\zero) \rangle}{\rnorm} =\frac{1}{\rnorm}\frac{\partial}{\partial\T}\langle \ethree,g_\T(\zero) \rangle+z\cdot\frac{\partial \|\mathbf{r}\|^{-1}}{\partial \mathbf{T}} \\
		=&\frac{1}{\rnorm}\left(\begin{bmatrix}
			\ethree^\top\mathbf{R} & \mathbf{0}_{1\times 3}
		\end{bmatrix}-\frac{z}{\rnorm^2}\cdot\begin{bmatrix}
			\mathbf{r}^\top \mathbf{R} & \mathbf{0}_{1\times 3}
		\end{bmatrix}\right).
	\end{split}
\end{equation}
Therefore, we have
\begin{equation}
	\begin{split}
		\J_\T^\theta &= -\frac{1}{\sqrt{1-(z/\rnorm)^2}}\cdot\frac{\partial}{\partial\T}\frac{\langle \ethree,g_\T(\zero) \rangle}{\rnorm} \\
		&=-\frac{1}{\sqrt{\rnorm^2-z^2}}\begin{bmatrix}
			\ethree^\top\mathbf{R}-\frac{z\cdot\mathbf{r}^\top \mathbf{R} }{\rnorm^2} & \mathbf{0}_{1\times 3}
		\end{bmatrix}.
	\end{split}
\end{equation}
\paragraph{$\mathbf{J}^{\phi}_{\mathbf{T}}$}
Considering (\ref{eq.1.38}), the dependence of delay $\phi$ on $\mathbf{T}$ is explicitly denoted as
\begin{equation}
	\begin{split}
		\phi&=\operatorname{atan2}\left(y,x\right)=\operatorname{atan2}(\langle \etwo,\rvec \rangle,\langle \eone,\rvec \rangle)\\
		&=\operatorname{atan2}(\langle \etwo,g_\T(\zero) \rangle,\langle \eone,g_\T(\zero) \rangle),
	\end{split}
\end{equation}
where $\etwo=[0,1,0]^\top\in\R^3$  and $\eone=[1,0,0]^\top\in\R^3$ are used to extract the element $y$ and $x$ from $\rvec$, respectively. Therefore, we have
\begin{align}
	\J_{\T}^{\phi}&=\frac{\partial \phi}{\partial x}\cdot \frac{\partial x}{\partial \T}+\frac{\partial \phi}{\partial y}\cdot \frac{\partial y}{\partial \T}  \nonumber \\
	&=\frac{-y}{x^2+y^2}\begin{bmatrix}
		\eone^\top\mathbf{R} & \zero_{1\times 3}
	\end{bmatrix}+\frac{x}{x^2+y^2}\begin{bmatrix}
		\etwo^\top\mathbf{R} & \zero_{1\times 3}
	\end{bmatrix} \nonumber\\
	&=\frac{1}{x^2+y^2}\begin{bmatrix}
		x\cdot\etwo^\top\mathbf{R}-y\cdot\eone^\top\mathbf{R} & \zero_{1\times 3}
	\end{bmatrix}.
\end{align}

\section{Appendix B}
\subsection{Trajectory Optimization Formulation as $(\mathrm{P}_2)$} \label{sub.1.2}
The subscript $n$ is omitted in the following for simplicity. According to \eqref{eq.1.72} and \eqref{eq.1.60}, $\mathbf{\Psi}\in\R^{4\times 6}$ can be partioned into
\begin{equation}
	\resizebox{.88\hsize}{!}{$
	\mathbf{\Psi}=\begin{bmatrix}
		\B & \zero \\
		\conet & \ctwot
	\end{bmatrix}=\begin{bmatrix}
	\B & \zero \\
	\coneonet & \ctwot
	\end{bmatrix}+\begin{bmatrix}
	\zero & \zero \\
	\conetwot & \zero
	\end{bmatrix}\triangleq\mathbf{\Psi}_1+\mathbf{\Psi}_2,
$}
\end{equation}
with
\begin{align}
  \coneonet &= \frac{2f_c}{c\rnorm}\cdot {^{\mathcal{B}}{\mathbf{v}}_{p}^\top}\mathbf{R}^\top\Prmatperp\mathbf{R}, \\
  \conetwot &= -\frac{2f_c}{c\rnorm}\cdot {^{\mathcal{B}}{\mathbf{v}}_{s}^\top}\Prmatperp\mathbf{R}, \\
  \mathbf{c}_2^\top &= -\frac{2f_c}{c\rnorm}\cdot \rvec^\top\mathbf{R}\left[{^{\mathcal{B}}{\mathbf{v}}_{p}}\right]_{\times}, 
\end{align}
and $\B\in\R^{3\times 3}$ is the remaining non-zero block. $\Prmatperp\triangleq\I- \rvec\cdot\rvec^\top / \rnorm^{2}$ is a projector onto the orthogonal complement of the subspace spanned by $\rvec$. The term $\conetwot$ is isolated as the only component that depends on ${^{\mathcal{B}}\vs^{n+1}}$ via ${^{\mathcal{B}}{\mathbf{v}}_{s}}$, which implies that $\rank(\mathbf{\Psi}_2)=1$ if $\conetwo\neq\zero$. Therefore, 
\begin{equation}
  \mathbf{\Psi}\mathbf{E}^{-1}\mathbf{\Psi}^\top = \sum_{i=1}^{2}\sum_{j=1}^{2}\mathbf{\Psi}_i\mathbf{E}^{-1}\mathbf{\Psi}_j^{\top}=\mathbf{D}_0+\mathbf{D},
\end{equation}
where $\mathbf{D}_0\triangleq\mathbf{\Psi}_1\mathbf{E}^{-1}\mathbf{\Psi}_1^{\top}\succcurlyeq\zero$, and $\mathbf{D}\triangleq\mathbf{D}_1+\mathbf{D}_2\triangleq\mathbf{\Psi}_2\mathbf{E}^{-1}(\mathbf{\Psi}_1^{\top}+\mathbf{\Psi}_2^{\top})+\mathbf{\Psi}_1\mathbf{E}^{-1}\mathbf{\Psi}_2^{\top}$ is a symmetric matrix. Since each term has rank $1$ due to $\rank(\mathbf{\Psi}_2)=1$, we have $\rank(\mathbf{D}) \leq 1 + 1 = 2$, and  $\rank(\mathbf{D}) = 0 \text{ or }2$. The matrix $\Dmat$ reads explicitly as
\begin{equation}
	\mathbf{D} = \begin{bmatrix}
		\mathbf{0} & \mathbf{d} \\
		\mathbf{d}^\top & d_0
	\end{bmatrix}, \quad
  \mathbf{E}^{-1} = \begin{bmatrix}
    \mathbf{E}_{11} & \mathbf{E}_{12} \\
    \mathbf{E}_{12}^\top & \mathbf{E}_{22}
	\end{bmatrix},
\end{equation}
where $\mathbf{d} = \mathbf{B}\mathbf{E}_{11}\mathbf{c}_{12}$, $d_0 = 2\mathbf{c}_{11}^\top\mathbf{E}_{11}\mathbf{c}_{12}+2\mathbf{c}_{2}^\top\mathbf{E}_{21}\mathbf{c}_{12} \nonumber+\mathbf{c}_{12}^\top\mathbf{E}_{11}\mathbf{c}_{12}$. If $\mathbf{c}_{12}\neq\zero$, $\mathbf{D}$ is indefinite and has rank $2$ with its inertia as $(1,1,2)$, which can be proven via the characteristic equation $\det(\mathbf{D}-\lambda\mathbf{I})=(-\lambda)^3(d_0-\lambda+\|\mathbf{d}\|^2 / \lambda)=0$, implying one positive and one negative eigenvalue. By rewritting $\mathbf{\Psi}_n\mathbf{E}^{-1}\mathbf{\Psi}_n^\top$, we have
\begin{equation} \label{eq.1.78}
  \mathrm{P_{2}}: \quad \min_{\vs^{n+1}} -\logdet\left(\tilde{\mathbf{A}}^{-1}+\mathbf{D}_0+\mathbf{D}\right) \quad \mathrm{s.t.} \quad (\ref{eq.1.86}),
\end{equation}
where $\mathbf{D} = \mathbf{e}_4\vs^{n+1,\top}\mathbf{K}^\top+\mathbf{K}\vs^{n+1}\mathbf{e}_4^\top+\vs^{n+1,\top}\mathbf{\tilde{\mathbf{P}}}\vs^{n+1}\mathbf{e}_4\mathbf{e}_4^\top$, $\Kmat \triangleq \Etilde^\top\Mmat\Smat$, and $\Ptilde \triangleq \Smat^\top\Mmat^\top\mathbf{E}_{11}\Mmat\Smat\succcurlyeq\zero$, and $\Etilde \triangleq \begin{bmatrix} \mathbf{E}_{11}\mathbf{B}^\top & \mathbf{E}_{11}\mathbf{c}_{11} + \mathbf{E}_{12}\mathbf{c}_{2} \\ \end{bmatrix}$, $\Mmat \triangleq -\frac{2f_c}{c\|\rvec\|}\cdot \mathbf{R}^\top\Prmatperp$, $    \Smat \triangleq \begin{bmatrix} -\left[\mathbf{s}\right]_\times & \I \\ \end{bmatrix}$. 

\subsection{Trajectory Optimization Formulation as $(\mathrm{P}_2^\prime)$} \label{sub.1.3}
Considering the sparsity of $\mathbf{D}$, the dimension of the problem can be further reduced by leveraging the determinant property of block matrices. By factorising $\tilde{\mathbf{A}}^{-1}+\mathbf{D}_0\succ\zero$, we have
\begin{equation} \label{eq.1.81}
  \tilde{\mathbf{A}}^{-1}+\mathbf{D}_0 \triangleq \begin{bmatrix}
    \Abar & \mathbf{a}\\
    \mathbf{a}^\top & a_0
  \end{bmatrix},
\end{equation}
whose determinant is given as $\det(\tilde{\mathbf{A}}^{-1}+\mathbf{D}_0)=\det(\Abar)\cdot(a_0-\mathbf{a}^\top\Abar^{-1}\mathbf{a})$. Similarly, we have $\det(\tilde{\mathbf{A}}^{-1}+\mathbf{D}_0+\mathbf{D})=\det(\Abar)\cdot\left((d_0+a_0)-(\mathbf{d}^\top+\mathbf{a}^\top)\Abar^{-1}(\mathbf{d}+\mathbf{a})\right)$, from which $\det(\tilde{\mathbf{A}}^{-1}+\mathbf{D}_0)$ can be separated, i.e,
\begin{align}
  &\det(\tilde{\mathbf{A}}^{-1}+\mathbf{D}_0+\mathbf{D}) \\
  =&\det(\tilde{\mathbf{A}}^{-1}+\mathbf{D}_0) + \det(\Abar) \cdot\left(d_0-2\mathbf{a}^\top\Abar^{-1}\mathbf{d}-\mathbf{d}^{\top}\Abar^{-1}\mathbf{d}\right) \nonumber \\
  =&\det(\tilde{\mathbf{A}}^{-1}+\mathbf{D}_0) \left(1+c_0\left(\mathbf{c}^\top\boldsymbol{\xi}_s^{n+1}+\vs^{n+1,\top}\mathbf{\bar{\mathbf{P}}}\vs^{n+1}\right)\right), \nonumber
\end{align}
with
\begin{align}
    c_0&\triangleq\det(\Abar) / \det(\tilde{\mathbf{A}}^{-1}+\mathbf{D}_0)>0, \\
    \mathbf{c}&\triangleq \Smat^\top\Mmat^\top\left(\mathbf{E}_{11}\coneone+\mathbf{E}_{12}\ctwo-\mathbf{E}_{11}\Bmat^\top\Abar^{-1}\mathbf{a}\right), \\
    \bar{\mathbf{P}}&\triangleq\tilde{\mathbf{P}}-\Smat^\top\Mmat^\top\mathbf{E}_{11}\Bmat^\top\Abar^{-1}\Bmat\mathbf{E}_{11}\Mmat\Smat \label{eq.1.83} \\
    &=\Smat^\top\Mmat^\top\mathbf{E}_{11}^{\frac{1}{2}}\left(\mathbf{I}-\mathbf{E}_{11}^{\frac{1}{2}}\Bmat^\top\bar{\mathbf{A}}^{-1}\Bmat\mathbf{E}_{11}^{\frac{1}{2}}\right)\mathbf{E}_{11}^{\frac{1}{2}}\Mmat\Smat. \nonumber
\end{align} 
By substituting \eqref{eq.1.81}-\eqref{eq.1.83} into \eqref{eq.1.78}, we have
\begin{equation}
  \min_{\vs^{n+1}} \ -\log\left(1+c_0\left(\vs^{n+1,\top}\mathbf{\bar{\mathbf{P}}}\vs^{n+1}+2\mathbf{c}^\top\boldsymbol{\xi}_s^{n+1}\right)\right).
\end{equation}
Since $\tilde{\mathbf{A}}^{-1}+\mathbf{D}_0+\mathbf{D}\succ\zero$, and $\tilde{\mathbf{A}}^{-1}+\mathbf{D}_0\succ\zero$,  the logarithm's argument is positive. Thus, the objective function becomes maximizing $\vs^{n+1,\top}\mathbf{\bar{\mathbf{P}}}\vs^{n+1}+2\mathbf{c}^\top\boldsymbol{\xi}_s^{n+1}$, i.e.,
\begin{equation}
  \mathrm{P_{2}^\prime}: \quad \max_{\vs^{n+1}} \ \vs^{n+1,\top}\mathbf{\bar{\mathbf{P}}}\vs^{n+1}+2\mathbf{c}^\top\boldsymbol{\xi}_s^{n+1}, \quad \mathrm{s.t.} \quad (\ref{eq.1.86}).
\end{equation}

\bibliographystyle{IEEEtran}
\bibliography{references}

%
%
%
%
%
%
%

\end{document}